\documentclass[11pt]{article}
\usepackage{amssymb,epsf,graphicx}
\usepackage{amsmath}
\allowdisplaybreaks[1]

\newcommand{\be}{\begin{equation}}
\newcommand{\ee}{\end{equation}}
\newcommand{\sectiono}[1]{\section{#1}\setcounter{equation}{0}}

\newcommand\crbig{\\\noalign{\vspace {1.5mm}}}

\def\Re{\mathop{\mathrm{Re}}}

\newtheorem{conjecture}{Conjecture}

\def\quin#1#2#3#4#5{
  \begin{array}{c}{
      \begin{picture}(34,39)(0,-17)
    \put(18,0){\line(4,3){13}}
    \put(18,0){\line(-1,3){5}}
    \put(18,0){\line(-1,0){16}}
    \put(18,0){\line(-1,-3){5}}
    \put(18,0){\line(4,-3){13}}
    \put(0,1){\makebox{$#1$}}
    \put(15,-17){\makebox{$#2$}}
    \put(32,-11){\makebox{$#3$}}
    \put(32,8){\makebox{$#4$}}
    \put(15,14){\makebox{$#5$}}
      \end{picture}
    }
  \end{array}
}
\def\quart#1#2#3#4{
  \begin{array}{c}{
      \begin{picture}(25,32)(0,-13)
    \put(0,11){\line(1,-1){22}}
    \put(0,-11){\line(1,1){22}}
    \put(2,10){\makebox{$#1$}}
    \put(23,10){\makebox{$#2$}}
    \put(3,-14){\makebox{$#3$}}
    \put(23,-14){\makebox{$#4$}}
      \end{picture}
    }
  \end{array}
}
\def\quartcub#1#2#3#4#5#6{
  \begin{array}{c}{
      \begin{picture}(53,39)(0,-17)
    \put(18,0){\line(-1,0){16}}
    \put(18,0){\line(0,1){16}}
    \put(18,0){\line(0,-1){16}}
    \put(18,0){\line(1,0){24}}
    \put(42,0){\line(3,5){8}}
    \put(42,0){\line(3,-5){8}}
    \put(0,1){\makebox{$#1$}}
    \put(20,14){\makebox{$#2$}}
    \put(20,-17){\makebox{$#3$}}
    \put(51,14){\makebox{$#4$}}
    \put(51,-17){\makebox{$#5$}}
    \put(28,3){\makebox{$#6$}}
      \end{picture}
    }
  \end{array}
}
\def\cubcub#1#2#3#4#5{
  \begin{array}{c}{
      \begin{picture}(43,35)(0,-15)
    \put(8,0){\line(-3,5){8}}
    \put(8,0){\line(-3,-5){8}}
    \put(8,0){\line(1,0){24}}
    \put(32,0){\line(3,5){8}}
    \put(32,0){\line(3,-5){8}}
    \put(3,12){\makebox{$#2$}}
    \put(3,-15){\makebox{$#1$}}
    \put(41,12){\makebox{$#4$}}
    \put(41,-15){\makebox{$#3$}}
    \put(18,3){\makebox{$#5$}}
      \end{picture} 
    }
  \end{array}
}
\def\cubcubcub#1#2#3#4#5#6#7{
  \begin{array}{c}{
      \begin{picture}(59,39)(0,-17)
    \put(28,0){\line(-1,0){20}}
    \put(8,0){\line(-3,5){8}}
    \put(8,0){\line(-3,-5){8}}
    \put(28,0){\line(0,-1){16}}
    \put(28,0){\line(1,0){20}}
    \put(48,0){\line(3,5){8}}
    \put(48,0){\line(3,-5){8}}
    \put(2,14){\makebox{$#1$}}
    \put(2,-17){\makebox{$#2$}}
    \put(30,-17){\makebox{$#3$}}
    \put(57,14){\makebox{$#4$}}
    \put(57,-17){\makebox{$#5$}}
    \put(14,3){\makebox{$#6$}}
    \put(35,3){\makebox{$#7$}}
      \end{picture}
    }
  \end{array}
}

\begin{document}
{}~ \hfill\vbox{\hbox{SISSA 22/2007/EP}}\break
\vskip 2.1cm

\centerline{\large \bf Closed Bosonic String Field Theory at Quintic Order II:}
\centerline{\large \bf Marginal Deformations and Effective Potential}

\vspace*{8.0ex}

\centerline{\large \rm Nicolas Moeller}

\vspace*{6.0ex}

\centerline{\large \it International School for Advanced Studies (SISSA)}
\centerline{\large \it via Beirut 2-4,}
\centerline{\large \it 34014 Trieste, Italy} \vspace*{2.0ex}
\centerline{E-mail: {\tt moeller@sissa.it}}

\vspace*{6.0ex}

\centerline{\bf Abstract}
\bigskip

We verify that the dilaton together with one exactly marginal field,
form a moduli space of marginal deformations of closed bosonic string
field theory to polynomial order five. We use the results of this
successful check in order to find the best functional form of a fit of
quintic amplitudes. We then use this fit in order to accurately
compute the tachyon and dilaton effective potential in the limit of
infinite level. We observe that to order four, the effective potential
gives unexpectedly accurate results for the vacuum. We are thus led to
conjecture that the effective potential, to a given order, is a good
approximation to the whole potential including {\em all} interactions
from the vertices up to this order from the untruncated string
field. We then go on and compute the effective potential to order
five. We analyze its vacuum structure and find that it has several
saddle points, including the Yang-Zwiebach vacuum, but also a local
minimum. We discuss the possible physical meanings of these vacua.

\vfill \eject

\baselineskip=16pt

\tableofcontents


\sectiono{Introduction}
\label{s_intro}

The main goal of this paper is to continue the search for a
nonperturbative closed bosonic string vacuum. Although this search in
the context of closed bosonic string field theory (CSFT) \cite{CSFT}
originally started in \cite{Kostelecky:1990mi, Belo-Zwie, Belo}, an
important breakthrough came in a paper by Yang and Zwiebach
\cite{vacuum} where it was realized that the ghost dilaton must be
included in the string field in the universal basis. Using the
solution of the quartic CSFT vertex \cite{quartic}, they found a
nonperturbative vacuum, namely an extremum of the potential truncated
to order four. Through an argument based on the low-energy effective
action of the closed tachyon, dilaton and massless fields, they
conjectured that a CSFT vacuum must have zero action. In another paper
\cite{Yang:2005rw}, they proposed that this vacuum corresponds to
infinite string coupling and that the universe undergoes a big crunch
when the tachyon has rolled to it.

The Yang-Zwiebach vacuum was subsequently studied with more accuracy
in \cite{Moe-Yang}. The CSFT action was still truncated to quartic
order, but fields of level up to ten where included in the string
field. The potential value at the vacuum was seen to converge to
approximately $-0.050$ (in units where $\alpha'=2$). It was then
concluded that the quintic terms of the potential should be included
in order to test the vanishing potential conjecture.

The quintic term of the CSFT action was calculated in
\cite{quintic}. The solution is numerical, it gives the Strebel
differentials determining the local coordinates, everywhere in the
reduced moduli space. This is a complicated calculation, which could
fortunately be checked by verifying the flatness of the dilaton
potential to order five; but we devote one section of this paper to a
further check of this solution. Namely we will calculate the effective
potential of the dilaton and one exactly marginal field, to order
five. This is the direct extension of a calculation done in
\cite{marginal, dilaton} to order four. As expected, we find that the
effective potential is flat (within the uncertainty on the quintic
terms), thereby successfully checking the quintic contact term
solution.

As a level truncation analysis similar to the one done in
\cite{vacuum, Moe-Yang}, would require, at order five, many contact
terms that are still time-consuming to compute, we decided to focus
instead on the effective tachyon and dilaton potential. We are able to
integrate out massive fields up to level twelve; but in order to
obtain the exact terms in the effective potential (those found after
integrating out all levels) we must extrapolate the results to
infinite level. We find that the fits used until now in the literature
are unsatisfactory; we therefore spend a section looking for the best
possible functional form of a fit, and we find a simple expression
that gives good results when checking the flatness of the dilaton and
marginal effective potential. We then go on and use this fit for the
calculation of the effective tachyon and dilaton potential to order
five. We first spend some time studying this potential to order
four. This allows us to observe that the Yang-Zwiebach vacuum found
from the effective potential, matches very accurately the solution
found from the potential to quartic order with {\em all} interactions
from a string field at a given level. This is surprising because the
effective potential lacks most of the quartic contact terms which are
included in the full quartic potential. We turn this observation into
an approximate conjecture, essentially stating that this remains true
at higher order. To order five, this would imply that the effective
potential, which requires only the quintic contact terms $\kappa^2
V_{t^5}$, $\kappa^2 V_{t^4d}$, $\kappa^2 V_{t^3d^2}$, $\kappa^2
V_{t^2d^3}$, $\kappa^2 V_{td^4}$, and $\kappa^2 V_{d^5}$, is a good
approximation to the potential to order five (which contains many many
more contact terms).

From the analysis of the effective potential to order five, we find
that the Yang-Zwiebach vacuum still exists to this order, and is
shallower than to order four, giving evidence for the vanishing of the
potential at the vacuum. An advantage of the effective potential is
that it allows to check easily if a given extremum is a local minimum,
maximum, or saddle point. We find that the Yang-Zwiebach vacuum is a
saddle point. But at order five, we also find a local minimum. We
discuss these implications in the last section.

At last, we want to look at the usual level truncation of the
potential, as was done in \cite{vacuum, Moe-Yang} to order four. We
were able to compute only a few contact terms, namely those of total
level not greater than four. Surprisingly, we see that once we
introduce the term of level two, the Yang-Zwiebach vacuum is destroyed
(and does not reappear at level four). Although this should be checked
at higher level, we argue that the effective potential analysis should
be more reliable than the standard level truncation.

\paragraph{}
This paper is structured as follows: In Section \ref{s_marginal}, we
verify the flatness of the potential in the combined dilaton and
marginal directions. We use the data computed there in order to find a
good universal fit in Section \ref{s_fits}. We can then proceed to the
computation of the effective potential in Section
\ref{s_potential}. The level truncation analysis is done in Section
\ref{s_lev-trunc}, and the results are compared and discussed in
Section \ref{s_conclusions}, where some physical interpretations are
also discussed. The technical details of the calculations of quintic
contact terms are collected in Appendix \ref{appA}.

\sectiono{Combined dilaton and marginal deformations}
\label{s_marginal}

There are two objectives in this section: We want to test further our
computations of the quintic contact terms; and we want to verify that
the effective potential of the dilaton together with an exactly
marginal field, is flat. Our code that computes quintic contact terms
\cite{quintic} was already successfully checked by verifying the
flatness of the dilaton effective potential at order five. This showed
that the five-dilaton contact term $\kappa^2 V_{d^5}$ has been
computed correctly; our code was thus seen reliable at least for the
computation of contact terms of five identical states. Here we want to
extend this check to the computations of terms involving two different
kinds of states; this is in fact all that will be needed in the rest
of this paper, either for the tachyon and dilaton effective potential
which requires the contact terms of $n$ tachyons and $5-n$ dilatons,
or for the potential with quintic terms to level four, which requires
the contact terms of four tachyons and one massive field. The
computations of quintic terms of states not all equal, involve some
(not difficult but not completely trivial) combinatorics and also some
symmetry of the reduced moduli space. The technical details are
explained in Appendix \ref{appA}. Concretely, we will verify the
flatness of the effective potential of the ghost dilaton $d$ and an
exactly marginal field $a$. The dilaton is given by 
\be 
d |D\rangle = d \, (c_1 c_{-1} - \bar{c}_1 \bar{c}_{-1}) |0\rangle, 
\ee 
and the marginal field is 
\be 
a |A\rangle = a \, \alpha^X_{-1} \bar{\alpha}^X_{-1} c_1 \bar{c}_1 |0\rangle, 
\label{marginal} \ee 
where we have singled out one spacetime dimension $X$. Our analysis is
the direct extension, to order five, of the analysis made by Yang and
Zwiebach in \cite{dilaton}. There the authors showed that the contact
terms $\kappa^2 V_{a^4}$ and $\kappa^2 V_{a^2d^2}$ are canceled by the
contributions from cubic interactions. In this section we will
similarly show that the contact terms $\kappa^2 V_{a^4d}$ and
$\kappa^2 V_{a^2d^3}$ are canceled by the contributions from cubic and
quartic vertices.

We start with the effective term $\kappa^2 V_{a^4d}^\mathrm{eff}$ of
four marginals and one dilaton. We write it diagrammatically as
\be
-4! \, i \, \kappa^2 V_{a^4d}^\mathrm{eff}=\quin{d}{a}{a}{a}{a} + \sum_i 
\quartcub{d}{a}{a}{a}{a}{\phi_i} + 
\sum_i \quartcub{a}{a}{a}{d}{a}{\phi_i}.
\label{diagram1} \ee
The easiest way to understand the coefficient in the left-hand side is
to note that the right-hand side is an amplitude, and that to form an
amplitude from a term in the potential one should include the
combinatorial factor (here $4!$ is the number of ways to assign the
four marginals) and a $-i$ (we are in Minkowski space, all vertices
bring a factor $-i$ and the propagators bring a factor $i$).  The
internal fields $\phi_i$ are all the scalar fields, except for the
marginal field and dilaton. More explicitly we construct the
components of the closed string field $|\Phi\rangle = \sum_i \phi_i
|\Phi_i\rangle$ in the Siegel gauge, from open fields $\tilde{\cal
O}_j|0\rangle$ and $\tilde{\cal O}_k|0\rangle$ of same levels and
arbitrary ghost numbers, provided they add up to two.
\be
|\Phi_i\rangle = \left(\tilde{\cal O}_j \tilde{\cal O}^\star_k - 
\tilde{\cal O}^\star_j \tilde{\cal O}_k \right) |0\rangle,
\label{phii} \ee
where the $\star$-conjugation changes left-moving oscillators into
right-moving ones and vice-versa without changing their order. The
expression (\ref{phii}) is invariant under world-sheet parity ${\cal
P}$ defined by ${\cal P} \Phi = - \Phi^\star$; it is easy to see from
an argument similar to the one in \cite{vacuum}, that we can
consistently restrict the string field to have ${\cal P}$-eigenvalue
one. The open fields belong to the Hilbert space 
\be
\tilde{\cal H}_\mathrm{open} = \mathrm{Span} \left\{ 
\alpha^X_{-i_1} \ldots \alpha^X_{-i_p} L'_{-j_1} \ldots L'_{-j_q}
b_{-k_1} \ldots b_{-k_r} c_{-\ell_1} \ldots c_{-\ell_s} c_1 |0\rangle
\right\},
\ee
where 
\be
i_1 \geq i_2 \geq \ldots i_p \geq 1, \quad
j_1 \geq j_2 \geq \ldots j_q \geq 2, \quad
k_1 \geq \ldots k_r \geq 1, \quad \ell_1 \geq \ldots \ell_s \geq 1,
\ee
and the $L'_{-j}$ are matter Virasoro operators in the
$25$-dimensional space orthogonal to $X$. We can further restrict the
closed string field by noting that in the diagrams (\ref{diagram1})
and all other diagrams in this section, the components
$|\Phi_i\rangle$ must couple via a cubic vertex to $n$ marginal fields
and $3-n$ dilatons. These couplings are zero unless the numbers of
$\alpha$'s and the number of $\bar{\alpha}$'s in $|\Phi_i\rangle$ have
the same parity which must be opposite to the parity of the ghost
numbers of the open fields composing $|\Phi_i\rangle$. Moreover, since
a Virasoro $L'_{-j}$ with odd index $j$ can couple only to another
Virasoro of odd index, we must have an even number of odd-indexed
Virasoro's in $|\Phi_i\rangle$. With the above rules it is
straightforward to construct the closed string field needed in this
section. At level zero, we have only the tachyon $tc_1 \bar{c}_1
|0\rangle$, at level two we have the dilaton and marginal field, then
at levels $4$, $6$, $8$, $10$ and $12$ (the highest level considered
in this paper) we have respectively $7$, $11$, $92$, $188$ and $1016$
fields.

We can now continue the calculation of the effective term $\kappa^2
V_{a^4d}^\mathrm{eff}$. First we separate the amplitude
(\ref{diagram1}) into a contact term and a Feynman term
\be
\kappa^2 V_{a^4d}^\mathrm{eff} = \kappa^2 V_{a^4d} + {\cal C}_{a^4d},
\ee 
and we will focus on the Feynman term 
\be -4! \, i \, {\cal
C}_{a^4d} = \sum_i \quartcub{d}{a}{a}{a}{a}{\phi_i} + \sum_i
\quartcub{a}{a}{a}{d}{a}{\phi_i}.
\label{diag1} \ee
Since at each level greater than zero we have several scalar fields,
which are in general not normalized, the propagators in (\ref{diag1})
will be nondiagonal matrices. We emphasize that the sums in
(\ref{diag1}) would be really meaningful only if the fields were
orthogonal, but in our case they must be understood schematically
although their meaning remains clear. It will be very convenient to
express each of the Feynman diagram in terms of matrix
multiplication. We introduce the following
notations. $\tilde{A}_{\phi_i \phi_j}$ and $\tilde{A}_{\phi_i \phi_j
\phi_k}$ are vectors\footnote{We reserve the untilded symbols for the
universal Hilbert space when we calculate the tachyon and dilaton
effective potential in Section \ref{s_potential}.}, whose components
are given by the coupling constants
\begin{align}
\left(\tilde{A}_{\phi_i \phi_j}\right)_k &\equiv \left\{ 
\Phi_i, \Phi_j, \Phi_k \right\} 
\nonumber \\
\left(\tilde{A}_{\phi_i \phi_j \phi_k}\right)_h &\equiv 
\left\{ \Phi_i, \Phi_j, \Phi_k, \Phi_h \right\},
\label {Atdef} 
\end{align}
and $\tilde{P}$ is the zero-momentum propagator, a matrix given by
\be
\tilde{P} = -\tilde{M}^{-1} \quad \text{where} \quad 
\tilde{M}_{ij} = \langle \Phi_i| c_0^- Q_B| \Phi_j \rangle.
\label{prop} \ee
We can now simply translate (\ref{diag1}) into
\be
-4! \, i \, {\cal C}_{a^4 d} = -6 \, i\, \tilde{A}_{a^2d}^T 
\tilde{P} \tilde{A}_{a^2} - 
4 \, i \, \tilde{A}_{a^3}^T \tilde{P} \tilde{A}_{ad},
\ee
where the only nontriviality is to write the combinatorial weights
of each diagram. Note that the factors $(-i)$ in the right-hand side come
from two vertices ($(-i)^2$) and one propagator ($i$). We thus have
\be
{\cal C}_{a^4 d} = \frac{1}{4} \tilde{A}_{a^2d}^T \tilde{P} \tilde{A}_{a^2} + 
\frac{1}{6} \tilde{A}_{a^3}^T \tilde{P} \tilde{A}_{ad}.
\label{Ca4d} \ee
We emphasize that the expression (\ref{Ca4d}) is exact in the
infinite level limit, where all the vectors $\tilde{A}$ and the matrix
$\tilde{P}$ have infinite size. In the level truncation we restrict
the internal fields $\phi_i$ in the propagators to have level not
greater than, say $\ell$.  And we define ${\cal C}_{a^4d}(\ell)$ by the
expression (\ref{Ca4d}) where the matrix $\tilde{P}$ and vectors
$\tilde{A}$ are truncated to finite size, including only the indices
related to fields of level smaller than or equal to $\ell$. The same
convention will apply to all other amplitudes ${\cal C}(\ell)$ in this
paper.  For the way to compute the quartic terms $\tilde{A}_{\phi_i
\phi_j \phi_k}$ we refer the reader to \cite{quartic, marginal,
dilaton, Moe-Yang}. We have computed them here up to level twelve, the
values of ${\cal C}_{a^4d}(\ell)$ are shown in Table \ref{margCTable}.

The computation of ${\cal C}_{a^2 d^3}$ is done in the same way. This time 
we have three diagrams
\be
-12 i \, {\cal C}_{a^2d^3} = \sum_i 
\quartcub{d}{d}{d}{a}{a}{\phi_i} + 
\sum_i \quartcub{d}{d}{a}{d}{a}{\phi_i} + 
\sum_i \quartcub{d}{a}{a}{d}{d}{\phi_i},
\ee
from which we can write
\be
{\cal C}_{a^2 d^3} = \frac{1}{12} \tilde{A}_{d^3}^T \tilde{P} \tilde{A}_{a^2} + 
\frac{1}{2} \tilde{A}_{ad^2}^T \tilde{P} \tilde{A}_{ad} + 
\frac{1}{4} \tilde{A}_{a^2d}^T \tilde{P} \tilde{A}_{d^2}.
\ee
And we present the values ${\cal C}_{a^2 d^3}(\ell)$ in Table
\ref{margCTable}. For completeness we also compute ${\cal
C}_{d^5}$. This amplitude was already computed in \cite{quintic} to
level ten and already seen to convincingly cancel the contact term,
but we want to extend it here to level twelve so that the calculation
is complete, and also so that we have more data to test the fits in
Section \ref{s_fits}. Here there is only one diagram, namely
\be
-5! \, i \, {\cal C}_{d^5} = \sum_i 
\quartcub{d}{d}{d}{d}{d}{\phi_i} \quad \Rightarrow \quad 
{\cal C}_{d^5} = \frac{1}{12} \tilde{A}_{d^3}^T \tilde{P} \tilde{A}_{d^2}.
\ee
And we list the values of ${\cal C}_{d^5}(\ell)$ in Table \ref{margCTable}.
\begin{table}[!ht]
\begin{center}
\renewcommand\arraystretch{1.5}
\vskip 0.1in
\begin{tabular}{|c|c|c|c|}
  \hline
  & ${\cal C}_{a^4d}(\ell)$ & ${\cal C}_{a^2d^3}(\ell)$ & ${\cal C}_{d^5}(\ell)$ \\
  \hline \hline
  $\ell=0$ & $2.09955$ & $-1.85370$ & $0.401963$ \\
  \hline
  $\ell=4$ & $1.43546$ & $-1.65253$ & $0.362003$ \\
  \hline
  $\ell=6$ & $1.42224$ & $-1.50815$ & $0.325946$ \\
  \hline
  $\ell=8$ & $1.38644$ & $-1.47248$ & $0.316744$ \\
  \hline
  $\ell=10$ & $1.38545$ & $-1.45971$ & $0.311198$ \\
  \hline
  $\ell=12$ & $1.38004$ & $-1.45361$ & $0.309417$ \\
  \hline \hline
  $\ell=\infty$ & $1.3774$ & $-1.4457$ & $0.3063$ \\ 
  \hline \hline
  contact term & $-1.3779 \pm 0.0024$ & $1.4452 \pm 0.0053$ & $-0.3063 \pm 0.0016$ \\
  \hline
\end{tabular}
\caption{\footnotesize{The marginal amplitudes from Feynman diagrams
with internal fields up to level twelve, and their extrapolations from
the fit (\ref{goodfit}). In the last line we list the contact terms
whose computations are explained in Appendix \ref{appA}.}}
\label{margCTable}
\end{center}
\end{table}
We also write in this table the extrapolated values ${\cal
C}(\infty)$ calculated from the fit (\ref{goodfit}) which will
be explained in Section \ref{s_fits}. And in the last line we show the
contact terms calculated with the program described in
\cite{quintic}. We relegate the technical details of the contact terms
calculations to Appendix \ref{appA}. We see from Table
\ref{margCTable}, that the contact terms cancel the contributions from
the Feynman diagrams with an accuracy well within the error margins on
the contact terms. This is good evidence that, as we expected, the
effective potential of the exactly marginal field $a$ and the dilaton
$d$, is flat. It also shows that the quintic contact terms of two
different kinds of fields, are computed correctly. In fact the
accuracy of the cancellation even suggests that the error on the
quintic terms has been overestimated. This possibility was already
discussed in \cite{quintic}, but at present this is still the best
error estimates that we can do.

\sectiono{Level truncation fits}
\label{s_fits}

In this section, we want to find and motivate a good functional form
for a fit of closed string amplitudes ${\cal C}(\ell)$ as functions of
the level $\ell$. We start by remembering that in open string field
theory, computations to very high levels (typically 100) have been
done (see for example \cite{Taylor:2002fy}) and it turns out that fits
of the form
$$
{\cal C}^\mathrm{fit}_\mathrm{open}(\ell) = f_0 + \frac{f_1}{\ell} + 
\frac{f_2}{\ell^2} + \frac{f_3}{\ell^3} + \ldots + \frac{f_N}{\ell^N}
$$
perform very well. We emphasize that the next-to-leading term is of
order $\ell^{-1}$, as was shown from the BST algorithm
\cite{BST}. Some particular closed string field theory amplitudes,
like
$$
\cubcub{a}{a}{a}{a}{\phi_i} \quad \text{or} \quad \cubcub{t}{t}{t}{t}{\psi_i},
$$
where the propagating fields $\phi_i$ and $\psi_i$ are tensor products
of twist-even open fields of ghost number one, can be expressed in
terms of open string amplitudes. In these cases it was shown in
\cite{marginal} that the next-to-leading order of the fit is
$\ell^{-2}$. One might then suggest that closed string amplitudes
should be fitted with
\be
{\cal C}^\mathrm{fit}(\ell) = f_0 + \frac{f_2}{\ell^2} + \frac{f_3}{\ell^3} + 
\ldots + \frac{f_N}{\ell^N} .
\label{fit1} \ee
But it was found \cite{dilaton} that this fit doesn't perform well for
amplitudes that cannot be expressed in terms of open physical
amplitudes. Instead, fits of the form
\be
{\cal C}^\mathrm{fit}(\ell) = f_0 + \frac{f_1}{\ell^\gamma}
\label{fit2} \ee
seem to work better once the exponent $\gamma$ has been adjusted in
some way.  In particular the authors of \cite{dilaton} found that
$\gamma=2.7$ and $\gamma = 3.2$ for the fits of ${\cal C}_{a^2d^2}$
and ${\cal C}_{a^4}$ respectively give the expected values as $\ell
\rightarrow \infty$ (the ones that cancel the quartic contact terms).

One could go on and imagine many variants of the above fits, for
example by adding a term $\frac{f_2}{\ell^{2\gamma}}$ to (\ref{fit2})
etc... In order to argue what fits are better, we must take a look at
Table \ref{margCTable}.  The first thing that we emphasize is that we
will keep only the data points $\ell = 4$, $\ell=8$ and
$\ell=12$. Indeed we see for example in the first column of the table,
that the values for $\ell=4$ and $\ell=6$ are very similar, as well as
the values for $\ell=8$ and $\ell=10$. This is easy to understand.
Fields of level $4n+2$ are made of open fields of odd level $2n+1$;
but in open string field theory, the parity of level is very
important, indeed the twist symmetry implies that the open vertex can
couple only an even number of odd level fields of ghost number one
(this is why one can consistently set these fields to zero in the
nonperturbative open string vacuum for example). So the similarities
between levels $4n$ and $4n+2$ are just remnants of twist
symmetry. Were we to plot ${\cal C}(\ell)$ for all values of $\ell$,
we would obtain a rather stair-looking curve, while if we keep only
levels $4n$ (or $4n+2$) the curve is smoother and thus easier to
fit. At last we throw away the value at $\ell=0$ as the fits are
singular there\footnote{One could of course fix that singular behavior
by, for example, replacing $\ell$ by $\ell+\ell_0$ in (\ref{fit1}) or
(\ref{fit2}), but we observed that the resulting fits are not
improved.}.

The second observation that we can make on Table \ref{margCTable}, is
that the values of ${\cal C}(\ell)$ behave {\em monotonically} with
the level $\ell$. We will assume that this monotonicity is a feature
of all amplitudes and persists at high level. For definiteness, let us
now consider a ${\cal C}(\ell)$ which is monotonically
decreasing. This monotonicity imposes strong restrictions on a good
fit of ${\cal C}(\ell)$ because we want the value of the fit at $\ell
\rightarrow \infty$ to be better, i.e. {\em smaller}, than the last
data point. If the number of data points that we are fitting is
greater than the number of parameters in our fit, the fit will not go
exactly through the data points, and there is an unacceptable risk
that the fit at infinity will give a value larger than our best data
point. There are other restrictions; indeed, if we take the fit
(\ref{fit1}) and all three of our data points, keeping thus three fit
parameters $f_0$, $f_2$ and $f_3$, it might happen that $f_2$ and
$f_3$ have different signs, which would imply that the fit is not
monotonically decreasing and we might again end up with a fitted value
at infinity worse than the best data point. We will therefore choose a
fit of the form (\ref{fit2}).
 
But we experienced that if we use the three data points at $\ell=4$,
$\ell=8$ and $\ell=12$ to set $f_0$, $f_1$ and $\gamma$, the fits are
sometimes quite poor in the sense that the value of the fit at $\ell
\rightarrow \infty$ does not satisfactorily cancel the quintic contact
term. But in those cases, we also observed that the value of $\gamma$
chosen by the fit, is far away from $3$. Let us then try to set
$\gamma=3$ from the beginning
\be
{\cal C}^\mathrm{fit}(\ell) = f_0 + \frac{f_1}{\ell^3}
\label{thefit3} \ee
and use the data points at $\ell=8$ and $\ell=12$ to determine $f_0$
and $f_1$, we have then explicitly
\be
{\cal C}^\mathrm{fit}(\infty) = f_0 = \frac{1}{19} 
\left(27 \, {\cal C}(12) - 8 \, {\cal C}(8) \right).
\label{prefit}\ee
The values from this fit for the marginal amplitudes of Section
\ref{s_marginal}, are shown in Table \ref{margCTable}; they cancel the
contact terms with a striking precision. The fit (\ref{thefit3})
therefore seems to be excellent, except for the amplitudes mentioned
at the beginning of this section, those whose internal (propagating)
fields are tensor products of physical (i.e. ghost number one)
twist-even open fields, whose fit we know should rather be
\be
{\cal C}^\mathrm{fit}(\ell) = f_0 + \frac{f_1}{\ell^2}.
\label{thefit2} \ee
All in all, we conclude that a good fit of closed amplitudes ${\cal
C}(\ell)$, is (\ref{thefit2}) if the internal fields are tensor
products of open physical and twist-even fields, and (\ref{thefit3})
otherwise, and that we should keep only the maximum available levels
$L$ and $L-4$ in order to determine $f_0$ and $f_1$. We can thus
express ${\cal C}^\mathrm{fit}(\infty)=f_0$ explicitly in terms of
${\cal C}(L)$ and ${\cal C}(L-4)$, namely
\be
\boxed{\begin{array}{l}
{\cal C}^\mathrm{fit}(\infty) = \displaystyle{
\frac{L^\gamma \, {\cal C}(L) - (L-4)^\gamma \, {\cal C}(L-4)}
{L^\gamma - (L-4)^\gamma},}
\crbig
\text{where} \ \gamma = \left\{ \begin{array}{l} 2 \quad
\text{if internal fields are} \otimes \text{of open phys. twist-even fields}
\\
3 \quad \text{otherwise}
\end{array} \right.
\end{array}}
\label{goodfit} \ee

In order to test further the fit (\ref{goodfit}) we redo, to level
twelve, the calculation of quartic marginal deformations that was done
in \cite{marginal,dilaton}. The results are shown in Table
\ref{quarticmargCTable}.
\begin{table}[!ht]
\begin{center}
\renewcommand\arraystretch{1.5}
\vskip 0.1in
\begin{tabular}{|c|c|c|c|}
  \hline
  & ${\cal C}_{a^4}(\ell)$ & ${\cal C}_{a^2d^2}(\ell)$ & ${\cal C}_{d^4}(\ell)$ \\
  \hline \hline
  $\ell=8$ & $0.265827$ & $-0.483015$ & $0.115777$ \\
  \hline
  $\ell=10$ & $0.265827$ & $-0.469970$ & $0.108550$ \\
  \hline
  $\ell=12$ & $0.259977$ & $-0.465334$ & $0.108499$ \\
  \hline \hline
  $\ell=\infty$ & $0.2553$ & $-0.4579$ & $0.1054$ \\ 
  \hline \hline
  contact term & $-0.2560$ & $0.4571$ & $-0.1056$ \\
  \hline
\end{tabular}
\caption{\footnotesize{The quartic marginal amplitudes from Feynman
diagrams at levels $8$, $10$ and $12$, and their extrapolations from
the fit (\ref{goodfit}). The last line shows the contact terms.}}
\label{quarticmargCTable}
\end{center}
\end{table}
The fit projections for ${\cal C}_{a^2d^2}$ and ${\cal C}_{d^4}$,
cancel the contact terms with substantially more accuracy than the
fits \cite{dilaton} from level six data. This is especially
interesting in the case of ${\cal C}_{a^2d^2}$; had we fitted it with
(\ref{fit2}) and $\gamma = 5/2$ as was done in \cite{dilaton}, we
would have found ${\cal C}^\mathrm{fit}_{a^2d^2}(\infty) = -0.4553$, a
worse result than what we find with $\gamma=3$. The fit of ${\cal
C}_{a^4}$ is however a little poorer here than in \cite{marginal}
(where the projection was $0.2559$). Note that the propagator of this
amplitude only involves fields which are tensor products of open
physical twist-even fields (this can also be seen from the fact that
the values at levels $8$ and $10$ are the same), and we should
therefore take $\gamma=2$. The fact that the data to level six gives a
better answer than the data to level twelve with the same functional
form of fit (with $\gamma=2$) is probably accidental. Anyway,
had we used $\gamma=3$ we would have found ${\cal
C}^\mathrm{fit}_{a^4}(\infty) = 0.2575$, not as good as with
$\gamma=2$. This is thus good evidence that the choice of $\gamma$ in
(\ref{goodfit}) is right.

\sectiono{The effective potential}
\label{s_potential}

We are now ready to confidently calculate the effective tachyon and
dilaton potential to order five. Indeed we have shown that we can
trust the quintic contact terms computations needed, and we have a
good fit at hand to extrapolate the results to infinite level. We
start with the order four (where quintic computations are not needed),
which had already been calculated in \cite{vacuum} to level four, but
we are going to level twelve and extrapolating; we will see that to
this order, the effective potential provides unexpectedly accurate
results for the Yang-Zwiebach vacuum \cite{vacuum}. We will then
proceed to order five and discuss the local extrema of the potential.

\paragraph{}
We start by giving here a few definitions. The closed string field
$|\Psi\rangle = \sum_i \psi_i |\Psi_i\rangle$ is in the universal
Hilbert space, and is as described in \cite{vacuum,Moe-Yang}. We again
split contact term and Feynman contribution
\be
\kappa^2 V_{\psi_1\psi_2\ldots\psi_N}^\mathrm{eff} = 
\kappa^2 V_{\psi_1\psi_2\ldots\psi_N} + {\cal C}_{\psi_1\psi_2\ldots\psi_N}.
\ee
And we use the following notations; $A_{\psi_i \psi_j}$ and 
$A_{\psi_i \psi_j \psi_k}$ are vectors, whose components are given by
\begin{align}
\left(A_{\psi_i \psi_j}\right)_k &\equiv \left\{ \Psi_i, \Psi_j, \Psi_k \right\} 
\nonumber \\
\left(A_{\psi_i \psi_j \psi_k}\right)_h &\equiv 
\left\{ \Psi_i, \Psi_j, \Psi_k, \Psi_h \right\} ,
\label {Adef} \end{align}
and $B_{\psi_i}$ are matrices with components
\be
\left( B_{\psi_i} \right)_{jk} \equiv \left\{ \Psi_i, \Psi_j, \Psi_k \right\} .
\label {Bdef} \ee
Since the multilinear string functions are totally symmetric, $B_{\psi_i}$ 
are symmetric matrices; and it doesn't matter in which order the index fields 
of $A$ are written. At last $P$ is the zero-momentum propagator, a matrix given by
\be
P = -M^{-1} \quad \text{where} \quad 
M_{ij} = \langle \Psi_i| c_0^- Q_B| \Psi_j \rangle.
\label{prop2} \ee

\subsection{Order four}
\label{s_order4}

We calculate here the terms $\kappa^2 V^\mathrm{eff}_{d^n t^{4-n}}$ for 
$n=0,\ldots,4$. The manipulations are similar to those of Section \ref{s_marginal}. 
Since the Feynman diagrams involve only cubic vertices, only those with an 
even number of dilaton can be nonzero. For ${\cal C}_{t^4}$ we find
\be
{\cal C}_{t^4} = \frac{i}{4!} \sum_i  \cubcub{t}{t}{t}{t}{\psi_i}
= \frac{1}{8} A_{tt}^T P A_{tt},
\ee
where the internal fields $\psi_i$ are all the scalars except the tachyon and dilaton. 
And for ${\cal C}_{t^2d^2}$ we have
\be
{\cal C}_{t^2d^2} = \frac{i}{4} \left( \sum_i  \cubcub{t}{t}{d}{d}{\psi_i} + 
\sum_i  \cubcub{t}{d}{t}{d}{\psi_i} \right)
= \frac{1}{4} A_{tt}^T P A_{dd} + 
\frac{1}{2} A_{td}^T P A_{td}.
\ee
The results to level twelve and their extrapolations are shown in Table \ref{EffPot4CTable}.
\begin{table}[!ht]
\begin{center}
\renewcommand\arraystretch{1.5}
\vskip 0.1in
\begin{tabular}{|c|c|c|}
  \hline
  $\ell$ & ${\cal C}_{t^4}(\ell)$ & ${\cal C}_{t^2d^2}(\ell)$ \\
  \hline \hline
  $4$ & $-\frac{1896129}{4194304} \approx -0.452072$ & 
  $\frac{25329}{16384} \approx 1.54596$ \\
  \hline
  $6$ & $-\frac{1896129}{4194304} \approx -0.452072$ & 
  $\frac{19104841}{11943936} \approx 1.59954$ \\
  \hline
  $8$ & $-\frac{24710749}{50331648} \approx -0.490958$ & 
  $\frac{178516846189}{104485552128} \approx 1.70853$ \\
  \hline
  $10$ & $-\frac{24710749}{50331648} \approx -0.490958$ & 
  $\frac{179239681645}{104485552128} \approx 1.71545$ \\
  \hline
  $12$ & $-\frac{16280361760337731}{32499186133893120} \approx -0.500947$ & 
  $\frac{17898902809317331}{10282945612677120} \approx 1.74064$ \\
  \hline \hline
  $\infty$ & $-0.5089$ & $1.754$ \\ 
  \hline
\end{tabular}
\caption{\footnotesize{The Feynman contributions needed for the computation of 
the effective potential at order four.}}
\label{EffPot4CTable}
\end{center}
\end{table}
The Feynman contribution for the term $\kappa^2 V^\mathrm{eff}_{d^4}$ is not needed 
because we can use the dilaton theorem
\be
0 = \quart{d}{d}{d}{d} + 
\sum_i \cubcub{d}{d}{d}{d}{\psi_i}  + \cubcub{d}{d}{d}{d}{t} = 
- 4! \, i \, \kappa^2 V^\mathrm{eff}_{d^4} - 3 \, i \, 
\left\{D,D,T\right\}\left(\frac{1}{2}\right) \left\{T,D,D\right\},
\ee 
from which we deduce
\be
\kappa^2 V^\mathrm{eff}_{d^4} = -\frac{1}{16} \left\{D,D,T\right\}^2 = 
-\frac{729}{4096} \approx -0.1780.
\ee
We now just need the contact terms (see \cite{vacuum} for example)
\be
\kappa^2 V_{t^4} = -3.017, \quad 
\kappa^2 V_{t^3d} = 3.872, \quad
\kappa^2 V_{t^2d^2} = 1.368, \quad
\kappa^2 V_{td^3} = -0.9528.
\ee
All in all we have for the potential at order four
\be
\kappa^2 V^\mathrm{eff}_4 = -t^2  + \frac{6561}{4096} \, t^3 - \frac{27}{32} \, td^2 
-3.526 \, t^4 + 3.872 \, t^3 d + 3.122 \, t^2 d^2 - 
0.9528 \, td^3 - \frac{729}{4096} \, d^4.
\label{Veff4}\ee

In order to judge how well it captures the vacuum structure, we will
compare the results for the local extremum found in truncation scheme
$B$ of \cite{Moe-Yang} and the analog found with the effective
potential truncated to fields of level $L$, with $L=4, 6, 8, 10$. The
analogs of (\ref{Veff4}) with internal fields of levels not greater
than $L$ are
\begin{align}
\kappa^2 V^\mathrm{eff}_{4,4} = & -t^2  + \frac{6561}{4096} \, t^3 - \frac{27}{32} \, td^2 
-3.469 \, t^4 + 3.872 \, t^3 d + 2.914 \, t^2 d^2 - 
0.9528 \, td^3 - 0.1390 \, d^4 \nonumber \\
\kappa^2 V^\mathrm{eff}_{4,6} = & -t^2  + \frac{6561}{4096} \, t^3 - \frac{27}{32} \, td^2 
-3.469 \, t^4 + 3.872 \, t^3 d + 2.968 \, t^2 d^2 - 
0.9528 \, td^3 - 0.1673 \, d^4 \nonumber \\
\kappa^2 V^\mathrm{eff}_{4,8} = & -t^2  + \frac{6561}{4096} \, t^3 - \frac{27}{32} \, td^2 
-3.508 \, t^4 + 3.872 \, t^3 d + 3.077 \, t^2 d^2 - 
0.9528 \, td^3 - 0.1678 \, d^4 \nonumber \\
\kappa^2 V^\mathrm{eff}_{4,10} = & -t^2  + \frac{6561}{4096} \, t^3 - \frac{27}{32} \, td^2 
-3.508 \, t^4 + 3.872 \, t^3 d + 3.083 \, t^2 d^2 - 
0.9528 \, td^3 - 0.1750 \, d^4
\label{Vefftrunc} \end{align}
We show in Table \ref{compvac}, the value of the potential for the
vacuum found in truncation scheme $B$ \cite{Moe-Yang} at fields level
$L$, compared to the values of the extrema of the potentials
(\ref{Vefftrunc}).
\begin{table}[!ht]
\begin{center}
\renewcommand\arraystretch{1.5}
\vskip 0.1in
\begin{tabular}{|c||c|c|c|c|c|}
  \hline
  $L$ & $4$ & $6$ & $8$ & $10$ & $\infty$ \\
  \hline \hline
  value of $\kappa^2 V^\mathrm{eff}_{4,L}$ & $-0.05443$ & $-0.05415$ & $-0.05266$ & 
  $-0.05274$ & $-0.05234$ \\
  \hline
  value of $\kappa^2 V_{L,4L}$ in scheme $B$ & $-0.05442$ & $-0.0544$ & $-0.0514$ & 
  $-0.0513$ & $-0.050$ \\
  \hline
\end{tabular}
\caption{\footnotesize{Comparison of the values of the effective
potential and the full potential at the nonperturbative vacuum of
\cite{vacuum,Moe-Yang}. The last line was calculated in Section 3 of
\cite{Moe-Yang}.}}
\label{compvac}
\end{center}
\end{table}
We emphasize that only the value at $L=4$ of $\kappa^2 V_{L,4L}$ in
truncation scheme $B$, is exact. The other ones were obtained by
extrapolating the values of $\kappa^2 V_{L,M}$ to $M=4L$. And the
value at infinity was in turn extrapolated from the values of the last
line of Table \ref{compvac}. We see a striking similarity between the
values at fields level $L=4$ (the small mismatch is within the
relative expected error made on the quartic terms, which is about
$0.1\%$). Could these values be exactly equal (and the mismatch of the
others be due to extrapolation errors)? We shouldn't expect so.
Indeed if we wanted to calculate the effective potential from the
potential, by solving the equations of motion for all the massive
fields for fixed values of $t$ and $d$, and plug back into the
potential the resulting expressions of the massive fields as functions
of $t$ and $d$, we should obtain a nonpolynomial function of $t$ and
$d$. This function would agree with $\kappa^2 V^\mathrm{eff}_{4,4}$ to
order four, but we will have terms of higher order as well. Those will
lack the contact terms of course, but they will contain terms from
Feynman diagrams built with cubic and quartic vertices. It is instructive to compare the 
tachyon and dilaton vacuum expectation values. From the effective potential 
$V^\mathrm{eff}_{4,4}$ we find
\be
(t,d) = (0.3424, 0.4057),
\ee
while from $V_{4,16}$ in scheme $B$ we find
\be
(t,d) = (0.3265, 0.4349).
\ee
This rules out strict equality, but these two results are not that
different. We will thus interpret the numerical values in Table
\ref{compvac}, as evidence for the following approximate conjecture.
\begin{conjecture}
The effective tachyon and dilaton potential $\kappa^2
V^\mathrm{eff}_N$ to a given polynomial order $N$, captures with good
approximation the physics of the whole potential including vertices up
to order $N$ and with {\em all} interactions from the untruncated
string field.
\label{conj} \end{conjecture}
We emphasize that this is not a precise statement as we are only
stating an approximation. This is nevertheless a strong statement; it
implies in particular that at order five, we may only calculate the
contact terms $\kappa^2V_{t^5}$, $\kappa^2V_{t^4d}$,
$\kappa^2V_{t^3d^2}$, $\kappa^2V_{t^2d^3}$, $\kappa^2V_{td^4}$ and
$\kappa^2V_{d^5}$ necessary to form the effective potential, and that
we will have a good approximation of the vacuum structure of the
potential with all quintic contact terms (to fields level four there
are $252$ such terms, to level six there are $20,349$ of them! And
then we would still need to extrapolate to infinite level).

\paragraph{}
Before going to order five, we want to do one more thing at order
four. We want to find all extrema of the potential (\ref{Veff4}) and
check whether they are local maxima, minima, or saddle points. In
order to do this we will look at the eigenvalues $\lambda_1$ and
$\lambda_2$ of the matrix $S$ of second derivatives
\be
S = \kappa^2 \begin{pmatrix} 
\partial^2_t V^\mathrm{eff} & \partial_t\partial_d V^\mathrm{eff} \\
\partial_d\partial_t V^\mathrm{eff} & \partial^2_d V^\mathrm{eff} 
\end{pmatrix}.
\ee 
Keeping only the real nontrivial solutions (and throwing away those
which are very close to the origin and merely artifacts of truncation)
we find three extrema. The one corresponding to the Yang-Zwiebach 
vacuum is
\be 
(t,d) = (0.3348, 0.4005), \quad
\kappa^2 V^\mathrm{eff}_4 = -0.05234, \quad (\lambda_1,\lambda_2) =
(-2.192, 1.810).  
\label{YZ4} \ee 
We have one negative and one positive eigenvalue, this vacuum is
therefore a saddle point. This is interesting, it means that it cannot
be a true vacuum of the theory. In other words, the theory expanded at
this vacuum still has a tachyon (of mass squared $\lambda_1$). What
about the other two vacua?  We have one vacuum with a negative dilaton vev
\be 
(t,d) = (0.2497, -0.8229), \quad
\kappa^2 V^\mathrm{eff}_4 = -0.06062, \quad (\lambda_1,\lambda_2) =
(-4.236, 1.148),  
\label{saddle1} \ee 
which is again a saddle point. The third vacuum has a negative tachyon vev
\be 
(t,d) = (-0.1312, -0.4829), \quad
\kappa^2 V^\mathrm{eff}_4 = -0.003062, \quad (\lambda_1,\lambda_2) =
(-1.967, 0.3736),  
\label{saddle2} \ee
again a saddle point. But we notice that $t$ and $\lambda_2$ are
rather small, we interpret this as this point belonging to the family
of vacua generated by the dilaton deformations of the perturbative
vacuum; it is an artifact of truncation that we find only a finite
number of these vacua.

\subsection{Order five}
\label{s_order5}

We now compute the effective potential to order five, and in the light
of the last section we hope that it may give us a good insight into
the vacuum structure of the theory. We start by calculating the
Feynman contributions
\begin{align}
{\cal C}_{t^5} =& \frac{i}{5!} \left( \sum_i \quartcub{t}{t}{t}{t}{t}{\psi_i} + 
\sum_{i,j} \cubcubcub{t}{t}{t}{t}{t}{\psi_i}{\psi_j} \right) 
= \frac{1}{12} A_{ttt}^T P A_{tt} + \frac{1}{8} A_{tt}^T P B_t P A_{tt}
\\
{\cal C}_{t^4d} =& \frac{i}{24} \sum_i \left( \quartcub{d}{t}{t}{t}{t}{\psi_i} + 
\quartcub{t}{t}{t}{d}{t}{\psi_i}\right)
= \frac{1}{4} A_{ttd}^T P A_{tt} + \frac{1}{6} A_{ttt}^T P A_{td}
\\
{\cal C}_{t^3d^2} =& \frac{i}{12} \sum_i \left(
\quartcub{d}{d}{t}{t}{t}{\psi_i} + \quartcub{d}{t}{t}{d}{t}{\psi_i} + 
\quartcub{t}{t}{t}{d}{d}{\psi_i} \right) \nonumber \\
& + \frac{i}{12} \sum_{ij} \left( \cubcubcub{d}{d}{t}{t}{t}{\psi_i}{\psi_j} + 
\cubcubcub{d}{t}{t}{d}{t}{\psi_i}{\psi_j} + 
\cubcubcub{d}{t}{d}{t}{t}{\psi_i}{\psi_j} \right) \nonumber \\
=& \frac{1}{4} A_{tdd}^T P A_{tt} + \frac{1}{2} A_{ttd}^T P A_{td} 
+ \frac{1}{12} A_{ttt}^T P A_{dd} \nonumber \\
& + \frac{1}{4} A_{dd}^T P B_t P A_{tt}
+ \frac{1}{2} A_{td}^T P B_t P A_{td}
+ \frac{1}{2} A_{td}^T P B_d P A_{tt}
\\
{\cal C}_{t^2d^3} =& \frac{i}{12} \sum_i \left(
\quartcub{t}{t}{d}{d}{d}{\psi_i} + \quartcub{t}{d}{d}{t}{d}{\psi_i} + 
\quartcub{d}{d}{d}{t}{t}{\psi_i} \right) \nonumber \\
&= \frac{1}{4} A_{ttd}^T P A_{dd} + \frac{1}{2} A_{tdd}^T P A_{td} 
+ \frac{1}{12} A_{ddd}^T P A_{tt}
\\
{\cal C}_{td^4} =& \frac{i}{24} \sum_i \left( \quartcub{t}{d}{d}{d}{d}{\psi_i} + 
\quartcub{d}{d}{d}{t}{d}{\psi_i}\right) + 
\frac{i}{24} \sum_{ij} \left( \cubcubcub{t}{d}{d}{d}{d}{\psi_i}{\psi_j} + 
\cubcubcub{d}{d}{t}{d}{d}{\psi_i}{\psi_j} \right) \nonumber \\
&= \frac{1}{4} A_{tdd}^T P A_{dd} + \frac{1}{6} A_{ddd}^T P A_{td} + 
\frac{1}{2} A_{td}^T P B_d P A_{dd} + \frac{1}{8} A_{dd}^T P B_t P A_{dd}.
\end{align}
The results are shown in Table \ref{EffPot5Table}.
\begin{table}[!ht]
\begin{center}
\renewcommand\arraystretch{1.5}
\vskip 0.1in
\begin{tabular}{|c|c|c|c|c|c|}
  \hline
  $\ell$ & ${\cal C}_{t^5}(\ell)$ & ${\cal C}_{t^4d}(\ell)$ & 
  ${\cal C}_{t^3d^2}(\ell)$ & ${\cal C}_{t^2d^3}(\ell)$ & 
  ${\cal C}_{td^4}(\ell)$ \\
  \hline \hline
  $4$ & $3.79575$ & $-1.55833$ & $-7.51218$ & $3.17206$ & $1.05369$ \\
  \hline
  $6$ & $3.79575$ & $-1.61549$ & $-8.15761$ & $3.41664$ & $1.33655$ \\
  \hline
  $8$ & $4.17801$ & $-1.73714$ & $-8.80564$ & $3.59308$ & $1.54958$ \\
  \hline
  $10$ & $4.17801$ & $-1.74333$ & $-8.89440$ & $3.61552$ & $1.62033$ \\
  \hline
  $12$ & $4.27270$ & $-1.77456$ & $-9.03854$ & $3.65374$ & $1.66610$ \\
  \hline \hline
  $\infty$ & $4.348$ & $-1.790$ & $-9.137$ & $3.679$ & $1.715$ \\ 
  \hline
\end{tabular}
\caption{\footnotesize{The Feynman contributions to the order five of
the effective potential, and their extrapolations to infinite level
using the fit (\ref{goodfit}).}}
\label{EffPot5Table}
\end{center}
\end{table}
\begin{table}[!ht]
\begin{center}
\renewcommand\arraystretch{1.5}
\vskip 0.1in
\begin{tabular}{|c|c|c|c|c|}
  \hline
  $\kappa^2 V_{t^5}$ & $\kappa^2 V_{t^4d}$ & $\kappa^2 V_{t^3d^2}$ & 
  $\kappa^2 V_{t^2d^3}$ & $\kappa^2 V_{td^4}$ \\
  \hline \hline
  $9.924 \pm 0.008$ & $-20.613 \pm 0.026$ & $4.702 \pm 0.021$ & $6.769 \pm 0.021$ & 
  $-0.8077 \pm 0.0036$ \\
  \hline
\end{tabular}
\caption{\footnotesize{The quintic contact terms needed at the order
five of the effective potential. Details on their computation can be
found in Appendix \ref{appA}.}}
\label{EffPot5ContactTable}
\end{center}
\end{table}
The corresponding contact terms are computed with the program
described in \cite{quintic}, and shown in Table
\ref{EffPot5ContactTable}.  The details are explained in Appendix
\ref{appA}.
For the term $\kappa^2 V^\mathrm{eff}_{d^5}$ we can again use the dilaton theorem 
to write
\be
0 = \quin{d}{d}{d}{d}{d} + 
\sum_i \quartcub{d}{d}{d}{d}{d}{\psi_i}  + \quartcub{d}{d}{d}{d}{d}{t} 
= - 5! \, i \, \kappa^2 V^\mathrm{eff}_{d^5} - 10 \, i \, 
\left\{D,D,D,T\right\}\left(\frac{1}{2}\right) \left\{T,D,D\right\},
\ee
and thus 
\be
\kappa^2 V^\mathrm{eff}_{d^5} = -\frac{1}{24} \left\{D,D,D,T\right\} 
\left\{T,D,D\right\} = -0.4020.
\ee
And finally we can write down the effective potential at order five
\be
\boxed{ \begin{array}{c}
\displaystyle{\kappa^2 V^\mathrm{eff}_5} = \displaystyle{
-t^2  + \frac{6561}{4096} \, t^3 - \frac{27}{32} \, td^2 
-3.526 \, t^4 + 3.872 \, t^3 d + 3.122 \, t^2 d^2 - 
0.9528 \, td^3 - \frac{729}{4096} \, d^4} 
\crbig 
+ 14.27 \, t^5 - 22.40 \, t^4 d - 4.435 \, t^3 d^2 +  
10.45 \, t^2d^3 + 0.9073 \, td^4 - 0.4020 \, d^5 \end{array}}.
\ee

We can now do the same vacuum search as we did to order four. This
time we find five real nontrivial extrema. The one corresponding to
the Yang-Zwiebach vacuum is
\be
\boxed{(t,d) = (0.2105, 0.4582), \quad
\kappa^2 V^\mathrm{eff}_5 = -0.03322, \quad (\lambda_1,\lambda_2) =
(-2.311, 1.870)}.  
\label{YZ5} \ee
In addition to this one, we find three other saddle points
\begin{align}
(t,d)& = (0.2676, -0.1185),& 
\kappa^2 V^\mathrm{eff}_5& = -0.03662,& (\lambda_1,\lambda_2)& =
(-0.5878, 4.594)& \nonumber \\
(t,d)& = (0.9881, 0.8575),& 
\kappa^2 V^\mathrm{eff}_5& = 0.06579,& (\lambda_1,\lambda_2)& =
(-3.112, 82.48)& \nonumber \\
(t,d)& = (-0.4221, -0.5721),& 
\kappa^2 V^\mathrm{eff}_5& = -0.07998,& (\lambda_1,\lambda_2)& =
(-9.067, 2.848).&
\label{saddles} \end{align}
But we now have a {\em minimum}
\be
\boxed{(t,d) = (0.4907, 0.3978), \quad
\kappa^2 V^\mathrm{eff}_5 = -0.08245, \quad (\lambda_1,\lambda_2) =
(0.9509, 8.841)}.  
\label{min} \ee

Before we discuss these results in Section \ref{s_conclusions}, we try 
the usual level truncation scheme in the next section.

\sectiono{Usual level truncation}
\label{s_lev-trunc}

In this section we want to address the question of tachyon
condensation in the level truncation by looking for extrema of the
potential itself (not the effective potential). There are two main
approaches to level truncation, which were denoted schemes $A$ and $B$
respectively in \cite{Moe-Yang}. Here, the analog of scheme $A$ would
be to expand the string field to a large given level and include as
many cubic and quartic interactions as possible, we would then include
quintic interactions level by level. In scheme $B$, we would increase
the level of the string field step by step, and include {\em all} the
cubic, quartic and quintic interactions. In \cite{Moe-Yang} it was
seen that convergence is better in scheme $B$, but the computations of
all quartic interactions was a challenge that could be completely
achieved only to string field level four. Here the quintic term is, of
course, even more challenging. At level two, the result is essentially
included in the effective potential discussed in Section
\ref{s_potential}. At level four, we would need to include all quintic
terms up to total level twenty (a total of 252 terms); this is beyond
the scope of this work. We will therefore focus on scheme $A$ in this
section.

We will truncate the string field to level four, namely 
\begin{align}
|\Psi\rangle & = t \, c_1 \bar{c}_1|0\rangle + d \, (c_1 c_{-1} - 
\bar{c}_1 \bar{c}_{-1}) |0\rangle + f_1 \, c_{-1} \bar{c}_{-1}|0\rangle
+ f_2 \, L_{-2} c_1 \bar{L}_{-2} \bar{c}_1 |0\rangle \nonumber \\
& \quad + f_3 \, (L_{-2} c_1 \bar{c}_{-1} - \bar{L}_{-2} \bar{c}_1 c_{-1}) |0\rangle + 
g_1 \, (b_{-2} c_1 \bar{c}_{-2} \bar{c}_1 - 
\bar{b}_{-2} \bar{c}_1 c_{-2} c_1) |0\rangle,
\label{field4} \end{align}
and we will include all the cubic and quartic interactions, and the
quintic interactions at levels zero, two and four. We will therefore
need the quintic contact terms $\kappa_2 V_{t^5}$, $\kappa_2 V_{t^4d}$
and $\kappa_2 V_{t^3d^2}$ (see Table \ref{EffPot5ContactTable}) and
the terms $\kappa_2 V_{t^4f_1}$, $\kappa_2 V_{t^4f_2}$, $\kappa_2
V_{t^4f_3}$ and $\kappa_2 V_{t^4g_1}$ shown in Table \ref{T4MTable}.
\begin{table}[!ht]
\begin{center}
\renewcommand\arraystretch{1.5}
\vskip 0.1in
\begin{tabular}{|c|c|c|c|}
  \hline
  $\kappa^2 V_{t^4 f_1}$ & $\kappa^2 V_{t^4 f_2}$ & $\kappa^2 V_{t^4 f_3}$ & 
  $\kappa^2 V_{t^4 g_1}$ \\
  \hline \hline
  $0.4059 \pm 0.0046 $ & $244.98 \pm 0.48$ & $-50.43 \pm 0.10$ & $-3.9353 \pm 0.0068$ \\
  \hline
\end{tabular}
\caption{\footnotesize{The contact terms of four tachyons and one field of level four.}}
\label{T4MTable}
\end{center}
\end{table}
The details of these computations can be found in Appendix
\ref{appA}. The quintic potentials at each level are thus
\begin{align}
& \kappa^2 V_0^{(5)} = 9.924 \, t^5 \nonumber \\
& \kappa^2 V_2^{(5)} = -20.61 \, t^4 d \\
& \kappa^2 V_4^{(5)} = 4.702 \, t^3 d^2 + t^4 \left( 0.4059 \, 
f_1 + 245.0 \, f_2 - 50.43 \, f_3 - 3.935 \, g_1 \right). \nonumber
\end{align}
And the total potentials are 
\begin{align}
& \mathbb{V}_0^{(5)} = \mathbb{V}_{4,16}^{(4)} + V_0^{(5)} \nonumber \\
& \mathbb{V}_2^{(5)} = \mathbb{V}_0^{(5)} + V_2^{(5)} \label{VV5} \\
& \mathbb{V}_4^{(5)} = \mathbb{V}_2^{(5)} + V_4^{(5)}, \nonumber
\end{align}
where $\mathbb{V}_{4,16}^{(4)}$ contains all the quadratic, cubic, and
quartic terms of fields of level up to four (and thus contains
interactions of level up to sixteen). We now look for a minimum of
these potentials corresponding to the Yang-Zwiebach vacuum. In order
to do this, we solve numerically the equations with a start value (a
seed) corresponding to this vacuum. The results are shown in Table
\ref{levtruncA}.
\begin{table}[!ht]
\begin{center}
\renewcommand\arraystretch{1.5}
\vskip 0.1in
\begin{tabular}{|c||c|c|c|c|c|c|c|}
\hline
\hbox{Potential} & $t$ & $d$ & $f_1$ & $f_2$ & $f_3$ & $g_1$ & 
\hbox{Value} \\
\hline \hline
$\kappa^2 \mathbb{V}_{4,16}^{(4)}$ & $0.3265$ & $0.4349$ & $-0.1221$ & 
$-0.008973$ & $-0.03845$ & $-0.09332$ & $-0.05442$ \\
\hline
$\kappa^2 \mathbb{V}_0^{(5)}$ & $0.2600$ & $0.2373$ & $-0.04735$ & 
$-0.004174$ & $-0.01530$ & $-0.03555$ & $-0.03281$ \\
\hline
$\kappa^2 \mathbb{V}_2^{(5)}$ & $0.2423$ & $-0.3718$ & $-0.009011$ & 
$0.0001399$ & $-0.003029$ & $0.02344$ & $-0.03802$ \\
\hline
$\kappa^2 \mathbb{V}_4^{(5)}$ & $0.1588$ & $-0.6072$ & $-0.04073$ & 
$-0.0005148$ & $-0.01074$ & $0.03996$ & $-0.02629$ \\
\hline
\end{tabular}
\caption{\footnotesize{The extremum of the potential found in the
level truncation scheme $A$.}}
\label{levtruncA}
\end{center}
\end{table}
We see that this vacuum is destroyed after we include the term of
level two $V_2^{(5)}$. Instead, a local extremum is found at a {\em
negative} value of the dilaton. We have done the same calculation with
$\mathbb{V}_0^{(5)} = \mathbb{V}_{10,10}^{(4)} + V_0^{(5)}$,
i.e. using fields up to level ten and with cubic interactions up to
level $24$ and quartics interaction up to level ten; and we found
qualitatively the same results as in Table \ref{levtruncA}. So the
breakdown of the solution is really due to the quintic terms. We found
another extremum to the potential $\mathbb{V}_4^{(5)}$ of (\ref{VV5}),
namely
\be
(t,d) = (-0.2031, -0.5240), \quad
\kappa^2 \mathbb{V}_4^{(5)} = -0.01152.
\label{extr} \ee

It is important to note that none of the extrema, (\ref{extr}) or the
one in Table \ref{levtruncA}, correspond to any extremum of the
effective potential of Section \ref{s_potential}.

\sectiono{Conclusions and prospects}
\label{s_conclusions}

In this paper we have shown that we are able to correctly compute
quintic contact terms when the interacting fields are not all the
same. This was shown by verifying, to order five, that the dilaton and
one exactly marginal field form a moduli space of marginal
deformations. We then used this data to motivate a universal fit which
gives very good approximations for all the verifiable amplitudes that
we have computed. This fit was then used in the computation of the
tachyon and dilaton effective potential. At order four, we noticed
that the extrema from this effective potential were very close (more
than expected) to the extrema found from the potential with many
terms. We phrased this nice apparent property as a conjecture.

\paragraph{}
Since it is only an approximate statement, we will interpret
Conjecture \ref{conj} as a statement on {\em level truncation}. In
other words it tells us that when including the vertex of order $N$,
one should first include the terms $\kappa^2 V_{t^n d^{N-n}}$ which
will be the most important contributions, and then include all the
terms with level four fields, and so on. This is different from usual
truncation as, for example, some terms of level $2N$ are included
before some terms of level $4$. It would be interesting to check such
a truncation scheme in a different context, like tachyons on orbifolds
(see \cite{twisted} for example).

\paragraph{}
It is a little bit surprising that, at order five, the vacua found
from the effective potential do not agree with those found in the
usual level truncation scheme $A$. If we do believe Conjecture
\ref{conj}, we shall give more credence to the results from the
effective potential. This is especially reasonable since we went only
to level four in the usual truncation scheme. We will take this point
of view, and not discuss further the results from usual truncation,
except to say that it would of course be interesting to include terms
of higher levels.

\paragraph{}
Of all the saddle points found from the effective potential, only one
seems physically meaningful. Indeed the solutions (\ref{saddles}) have
no equivalent at order four; and similarly the saddle points
(\ref{saddle1}) and (\ref{saddle2}) have no analog at order five. The
Yang-Zwiebach vacuum (\ref{YZ4}), however, survives to order five;
moreover the eigenvalues $\lambda_i$ are stable from order four to
order five. This is evidence that this vacuum is physical, present in
the full untruncated theory. The value of the potential at this vacuum
goes from $-0.05234$ at order four to $-0.03322$ at order five. This
is certainly compatible with the conjecture \cite{vacuum} that it
should be zero. On the other hand, one might be concerned by the fact
that the vacuum expectation value of the tachyon goes from $0.3348$ at
order four to $0.2105$ at order five. Is this vacuum simply going to
converge to a dilaton deformation of the perturbative vacuum to higher
order? One of the eigenvalues $\lambda_i$ should then tend to zero,
but this is clearly not the case, as can be seen from (\ref{YZ4}) and
(\ref{YZ5}). We are thus led to claim that this vacuum is physically
interesting. As to its interpretation, the shallowness of the
potential certainly supports the interpretation from the low-energy
effective action \cite{vacuum,Yang:2005rw} that the universe ends in a
big crunch there. But the fact that the Yang-Zwiebach vacuum is not a
local minimum but a saddle point certainly raises new questions. On
the one hand, one could argue that the big crunch interpretation is so
drastic that it doesn't matter that we are not on a stable point. It
is even tempting to imagine that the remaining instability could bring
the system back to its original perturbative vacuum, and that the
universe would thus undergo an infinite cycle of big crunches and big
bangs, like in cyclic universe models \cite{cyclic}. On the other
hand, one might wonder whether the system will ever reach the saddle
point. Indeed, even if the system starts rolling approximately towards
it, it seems natural that it will eventually turn to the downward
direction and miss it.

\paragraph{}
But in this paper we have found a local minimum as well
(\ref{min}), a very interesting result as it suggests the existence of
a stable nonperturbative vacuum. This is found only at order five and
has no analog at lower order, it is thus hard to say at this point
whether this is a physical result or just an effect of truncation. As
for its physical interpretation, it is as hard to say. We can
nevertheless note that it has a positive tachyon vev - what we naively
expect from a vacuum since negative tachyon values correspond to the
unbounded side of the potential at cubic order. It has also a positive
dilaton vev, corresponding to large string coupling as argued in
\cite{Yang:2005rw, vacuum}. Some clue could be given by the second
derivatives of the potentials (the eigenvalues $\lambda_1$ and
$\lambda_2$) which should correspond to the mass squared of two
particles found in this vacuum. Those are respectively approximately
$1$ and $9$ (in units where $\alpha'=2$). 

\paragraph{}
There are several directions in which the present work could be
continued. In particular, more quintic contact terms could be
computed. This could in particular allow to check Conjecture
\ref{conj}, and see if the Yang-Zwiebach vacuum is restored in the
usual level truncation after including more terms. If we want to
continue the direct search of a nonperturbative vacuum, however, it
seems very desirable to be able to make computations at order six. An
extension of \cite{quintic} to the sixtic term, however, would require
tremendous work and very strong programming skills. Other approaches
should be considered. Progress on the analytical side would be of
course extremely important, but a different numerical approach might
be the way to go. For example, if we remember that the most
complicated part in the contact term computation \cite{quintic} was
the computation of the boundary of the reduced moduli space, a natural
suggestion is to integrate over the {\em whole} moduli space
instead. We would thus produce effective terms (which is good if we
believe Conjecture \ref{conj}); but we would encounter divergences as
well, coming from the propagator of the zero-momentum dilaton. It
would therefore be very interesting to find a way to deal with these
divergences (Belopolsky managed to do this at order four \cite{Belo}).

\section*{Acknowledgments}
I thank N.~Berkovits for useful discussions, and H.~Yang and
B.~Zwiebach for comments on the manuscript. And I wish to thank the
organizers of the informal string theory workshop at HRI in Allahabad,
where part of this work was done, for their hospitality. This work has
been funded by an "EC" fellowship within the framework of the "Marie
Curie Research Training Network" Programme, Contract
no. MRTN-CT-2004-503369.

\appendix

\sectiono{Quintic contact terms}
\label{appA}

We collect here the technical results needed to compute the quintic
contact terms needed in this paper. All the closed string correlators
are given explicitly. Their integration over the moduli space was done
with the program developed in \cite{quintic}; for more details the
reader should consult this reference.

\subsection{Integration over the reduced moduli space}

We begin by recalling how to integrate over the reduced moduli space
of spheres with five punctures ${\cal V}_{0,5}$. It was shown in
\cite{quintic} that this space can be divided into 120 pieces and that
the integration can be written as an integration over one single piece
${\cal A}_5$. The five-string multilinear function reads
\be
\left\{\Psi_1, \Psi_2, \Psi_3, \Psi_4, \Psi_5 \right\} = 
\frac{1}{\pi^2} \int_{{\cal V}_{0,5}}dx_1 dy_1 dx_2 dy_2 
\langle \Sigma | \left( {\cal B} {\cal B}^\star \right)_1 
\left( {\cal B} {\cal B}^\star \right)_2 | \Psi_1 \rangle  
| \Psi_2 \rangle | \Psi_3 \rangle | \Psi_4 \rangle | \Psi_5 \rangle, 
\label{multilinear1} \ee
where the antighost insertions $\left({\cal B}{\cal B}^\star\right)_i$ are given by
\begin{align}
&{\cal B}_i = \sum_{I=1}^5 \sum_{m=-1}^\infty \left( B_{i,m}^I b_m^{(I)} + 
\overline{C_{i,m}^I} \bar{b}_m^{(I)} \right), \quad 
{\cal B}_i^\star = \sum_{I=1}^5 \sum_{m=-1}^\infty \left(C_{i,m}^I b_m^{(I)} + 
\overline{B_{i,m}^I} \bar{b}_m^{(I)} \right) \\
& B_{i,m}^{(I)} = \oint \frac{dw}{2\pi i} \frac{1}{w^{m+2}} \frac{1}{h_I'} 
\frac{\partial h_I}{\partial \xi_i} , \quad 
C_{i,m}^{(I)} = \oint \frac{dw}{2\pi i} \frac{1}{w^{m+2}} \frac{1}{h_I'} 
\frac{\partial h_I}{\partial \bar{\xi}_i},
\end{align}
with $h_I$ being the maps from the local coordinates $w_I$ at the
puncture $I$ to the uniformizer $z$ on the sphere
\be
z = h_I(w_I;\xi_1, \bar{\xi}_1, \xi_2, \bar{\xi}_2) = z_I + \rho_I w_I +
\rho_I^2 \beta_I w_I^2 + \rho_I^3 \gamma_I w_I^3 + {\cal O}(w_I^4).
\label{mapsI} \ee
All the coefficients in the right-hand side depend on the complex numbers 
$\xi_1=x_1+i y_1$ and $\xi_2=x_2+i y_2$ that parameterize the five-punctured spheres, 
and the $z_I$ are the punctures, where the states are inserted, $z_1=0$, $z_2=1$, 
$z_3=\xi_1$, $z_4=\xi_2$. The fifth puncture is at $z=\infty$ and there we 
should use the coordinate $t = 1/z$
\be
t = h_5(w_5;\xi_1, \bar{\xi}_1, \xi_2, \bar{\xi}_2) = \rho_5 w_5 +
\rho_5^2 \beta_5 w_5^2 + \rho_5^3 \gamma_5 w_5^3 + {\cal O}(w_5^4).
\label{maps5} \ee
All these coefficients can be expressed in terms of the quadratic differential 
defining the geometry of the punctured sphere, which can in turn be expressed 
numerically in terms of $\xi_1$ and $\xi_2$ (see \cite{quintic}).

In \cite{quintic}, the five states $|\Psi_i\rangle$ in (\ref{multilinear1}) 
where the same and the integral could simply be written as 120 times the integral 
of the same function over ${\cal A}_5$. We now want to deal with the case where 
the states $|\Psi_i\rangle$ are different. We start by defining
\be
F(\Psi_1,\Psi_2,\Psi_3|\Psi_4,\Psi_5) \equiv
\langle \Sigma | \left( {\cal B} {\cal B}^\star \right)_1 
\left( {\cal B} {\cal B}^\star \right)_2 | \Psi_1 \rangle  
| \Psi_2 \rangle | \Psi_4 \rangle | \Psi_5 \rangle | \Psi_3 \rangle .
\ee
Note how we have separated the states $\Psi_4$ and $\Psi_5$ from the other ones. 
These are inserted on the punctures $z=\xi_1$ and $z=\xi_2$ respectively. The construction 
of the reduced moduli space done in \cite{quintic} was such that these punctures always 
are on triangular faces of the interaction polyhedron, whereas the other three punctures 
$z=0$, $z=1$ and $z=\infty$ are always on quadrilateral faces. This is convenient because 
it makes visible the symmetry under the six $\mathrm{PSL}(2,\mathbb{Z})$ maps that permute the 
points $0$, $1$ and $\infty$. It will be convenient to explicitly name these maps
\be
s_1(z) = z , \quad s_2(z) = \frac{1}{z} , \quad s_3(z) = 1-z , 
\quad s_4(z) = \frac{1}{1-z} , \quad s_5(z) = \frac{z-1}{z} ,
\quad s_6(z) = \frac{z}{z-1} .
\ee
We can then write
\begin{align}
& \int_{{\cal V}_{0,5}} dx_1 dy_1 dx_2 dy_2 \langle \Sigma| 
\left( {\cal B} {\cal B}^\star \right)_1 \left( {\cal B} {\cal B}^\star \right)_2 
|\Psi_1\rangle |\Psi_2\rangle |\Psi_3\rangle |\Psi_4\rangle |\Psi_5\rangle  = 
\nonumber \\
& = \sum_{i=1}^6 \left( \int_{s_i({\cal A}_5)} +  
\int_{\overline{s_i({\cal A}_5)}} \right)
\Bigl(F(\Psi_1, \Psi_2, \Psi_3 | \Psi_4, \Psi_5) + \text{permutations} \Bigr) 
dx_1 dy_1 dx_2 dy_2 ,
\label{intV05}
\end{align} 
where the permutations denote the ten different ways of assigning
three states to the first three arguments of $F$ regardless of
order. In other words those are the ten different ways of assigning
three states to the quadrilateral faces. The integrals over the
complex conjugates $\overline{s_i({\cal A}_5)}$ can be easily related
to the integrals over $s_i({\cal A}_5)$ after we note that the
parameters $a_i$ of the quadratic differentials (see \cite{quintic})
obey $a_i(\overline{\xi_1}, \overline{\xi_2}) = \overline{a_i}(\xi_1,
\xi_2)$, $i=1,2$.  We simply have 
$$ 
\int_{\overline{s_i({\cal A}_5)}}
F(\Psi_1, \Psi_2, \Psi_3 | \Psi_4, \Psi_5)dx_1 dy_1 dx_2 dy_2 = 
\int_{s_i({\cal
A}_5)} \overline{F(\Psi_1, \Psi_2, \Psi_3 | \Psi_4,
\Psi_5)} dx_1 dy_1 dx_2 dy_2 .
$$
And since our states always obey the reality condition, we
have $F = \overline{F}$ on the Hilbert spaces we are
considering in this paper. Thus 
\be
\int_{\overline{s_i({\cal A}_5)}}
F(\Psi_1, \Psi_2, \Psi_3 | \Psi_4, \Psi_5)dx_1 dy_1 dx_2 dy_2 = 
\int_{s_i({\cal
A}_5)} F(\Psi_1, \Psi_2, \Psi_3 | \Psi_4,
\Psi_5) dx_1 dy_1 dx_2 dy_2 .
\label{reality} \ee
For (\ref{intV05}) to make sense we still need to show that the order of the first three 
arguments and the order of the last two 
arguments do not matter in the expression
$$
\sum_{i=1}^6 \int_{s_i({\cal A}_5)} F(\Psi_1, \Psi_2, \Psi_3 | \Psi_4, \Psi_5) 
dx_1 dy_1 dx_2 dy_2.
$$ 
To show that, we first remind that, in \cite{quintic}, we defined the
space ${\cal V}_{0,5}^{\{0,1,\infty\}}$ to be the subspace of ${\cal
V}_{0,5}$ for which the punctures at $0$, $1$ and $\infty$ are on
quadrilateral faces. It can be written
\be
{\cal V}_{0,5}^{\{0,1,\infty\}} = \bigcup_{i=1}^6 \left( s_i({\cal A}_5) 
\cup \overline{s_i({\cal A}_5)} \right) .
\label{Nu01inf} \ee
From its definition, this space is obviously symmetric under the exchange
$\xi_1 \leftrightarrow \xi_2$, which corresponds to the exchange of the last two
arguments of $F$. Therefore we have, using (\ref{reality}) and (\ref{Nu01inf})
\be
\sum_{i=1}^6 \int_{s_i({\cal A}_5)} F(\Psi_1, \Psi_2, \Psi_3 | \Psi_4, \Psi_5) 
dx_1 dy_1 dx_2 dy_2 =  
\sum_{i=1}^6 \int_{s_i({\cal A}_5)} F(\Psi_1, \Psi_2, \Psi_3 | \Psi_5, \Psi_4)
dx_1 dy_1 dx_2 dy_2 .
\label{perm2} \ee
At last, the integrations over $s_i({\cal A}_5)$ can be written as integrals 
over ${\cal A}_5$ after permutations of the first three punctures. Namely
\begin{align}
\int_{s_2({\cal A}_5)} F(\Psi_1, \Psi_2, \Psi_3 | \Psi_4, \Psi_5)
dx_1 dy_1 dx_2 dy_2 &= \int_{{\cal A}_5} F(\Psi_3, \Psi_2, \Psi_1 | 
\Psi_4, \Psi_5) dx_1 dy_1 dx_2 dy_2 \nonumber \\
\int_{s_3({\cal A}_5)} F(\Psi_1, \Psi_2, \Psi_3 | \Psi_4, \Psi_5) 
dx_1 dy_1 dx_2 dy_2 &= \int_{{\cal A}_5} F(\Psi_2, \Psi_1, \Psi_3 | 
\Psi_4, \Psi_5) dx_1 dy_1 dx_2 dy_2 \nonumber \\
\int_{s_4({\cal A}_5)} F(\Psi_1, \Psi_2, \Psi_3 | \Psi_4, \Psi_5) 
dx_1 dy_1 dx_2 dy_2 &= \int_{{\cal A}_5} F(\Psi_2, \Psi_3, \Psi_1 | 
\Psi_4, \Psi_5) dx_1 dy_1 dx_2 dy_2 \nonumber \\
\int_{s_5({\cal A}_5)} F(\Psi_1, \Psi_2, \Psi_3 | \Psi_4, \Psi_5) 
dx_1 dy_1 dx_2 dy_2 &= \int_{{\cal A}_5} F(\Psi_3, \Psi_1, \Psi_2 | 
\Psi_4, \Psi_5) dx_1 dy_1 dx_2 dy_2 \nonumber \\
\int_{s_6({\cal A}_5)} F(\Psi_1, \Psi_2, \Psi_3 | \Psi_4, \Psi_5) 
dx_1 dy_1 dx_2 dy_2 &= \int_{{\cal A}_5} F(\Psi_1, \Psi_3, \Psi_2 | 
\Psi_4, \Psi_5) dx_1 dy_1 dx_2 dy_2
\label{perm3} \end{align}
Now from (\ref{multilinear1}), (\ref{intV05}), (\ref{reality}), (\ref{perm2}) 
and (\ref{perm3}) we can simply write
\be
\boxed{\left\{\Psi_1, \Psi_2, \Psi_3, \Psi_4, \Psi_5 \right\} = 
\frac{1}{\pi^2} \sum_{\sigma \in S_5} \int_{{\cal A}_5} F(\Psi_{\sigma(1)}, 
\Psi_{\sigma(2)}, \Psi_{\sigma(3)} | \Psi_{\sigma(4)}, \Psi_{\sigma(5)})
dx_1 dy_1 dx_2 dy_2} ,
\label{intsym} \ee
where we sum over all the 120 elements of the permutation group $S_5$
of five elements. This is the most symmetric way of writing the
multilinear function as an integral over ${\cal A}_5$. We now want to
specialize this formula for the two special cases encountered in this
paper, when we have only two different kinds of states.

\paragraph{}
Now assume that we have $\Psi_1 = \ldots = \Psi_4 = \Phi$ and $\Psi_5 = \Psi$. From 
(\ref{intsym}) we can write
\begin{align}
\kappa^2 V_{\phi^4 \psi} = \frac{1}{24}\left\{ \Psi, \Phi, \Phi, \Phi, \Phi \right\} 
=& \frac{1}{\pi^2} \int_{{\cal A}_5} \Bigl( F(\Psi, \Phi, \Phi| \Phi, \Phi) 
+ F(\Phi, \Psi, \Phi| \Phi, \Phi) + F(\Phi, \Phi, \Psi| \Phi, \Phi) 
\nonumber \\
& + F(\Phi, \Phi, \Phi| \Psi, \Phi) + F(\Phi, \Phi, \Phi| \Phi, \Psi) 
\Bigr) dx_1 dy_1 dx_2 dy_2 .
\end{align}
From (\ref{perm2}) and (\ref{perm3}), we have that
\be
\int_{{\cal A}_5}F(\Phi, \Phi, \Phi| \Psi, \Phi) dx_1 dy_1 dx_2 dy_2 = 
\int_{{\cal A}_5}F(\Phi, \Phi, \Phi| \Phi, \Psi) dx_1 dy_1 dx_2 dy_2 ,
\ee
Introducing the definition 
\be
F_{\phi^4 \psi}^{(I)} \equiv \langle \Sigma| 
\left( {\cal B} {\cal B}^\star \right)_1 \left( {\cal B} {\cal B}^\star \right)_2 
|\Psi^{(I)} \Phi^{(J)} \Phi^{(K)} \Phi^{(L)} \Phi^{(H)} \rangle
\label{defI} \ee
where the state $\Psi$ is inserted on the puncture $I$, and the $\Phi$'s are 
inserted on the other four punctures $J$, $K$, $L$ and $H$. We can now write
\be
\boxed{\kappa^2 V_{\phi^4 \psi} = \frac{1}{\pi^2} \int_{{\cal A}_5} 
\left( F_{\phi^4 \psi}^{(1)} + F_{\phi^4 \psi}^{(2)} + F_{\phi^4 \psi}^{(5)} + 
2 F_{\phi^4 \psi}^{(3)} \right) dx_1 dy_1 dx_2 dy_2} .
\label{phi4psi} \ee
This is not the most symmetric way to write the amplitude, but it
involves less different functions $F_{\phi^4 \psi}^{(I)}$, which are
quite long expressions that take time to calculate.

\paragraph{}
Next we assume that $\Psi_1 = \ldots = \Psi_3 = \Phi$ and $\Psi_4 = \Psi_5 = \Psi$. 
This time we have 
\be
\kappa^2 V_{\phi^3 \psi^2} = \frac{1}{12}\left\{ \Psi, \Psi, \Phi, \Phi, \Phi \right\} .
\ee
Again we can use (\ref{perm2}) to reduce a little bit the number of different functions 
in the integral, noting that
\begin{align}
&\int_{{\cal A}_5}\left( F(\Psi, \Phi, \Phi| \Psi, \Phi) + 
F(\Phi, \Psi, \Phi| \Psi, \Phi) + F(\Phi, \Phi, \Psi| \Psi, \Phi) \right) 
dx_1 dy_1 dx_2 dy_2 = \nonumber \\
& = \int_{{\cal A}_5}\left(F(\Psi, \Phi, \Phi| \Phi, \Psi) + 
F(\Phi, \Psi, \Phi| \Phi, \Psi) + F(\Phi, \Phi, \Psi| \Phi, \Psi) \right) 
dx_1 dy_1 dx_2 dy_2 .
\nonumber \end{align}
Extending the definition (\ref{defI}) with
\be
F_{\phi^3 \psi^2}^{(IJ)} \equiv \langle \Sigma| 
\left( {\cal B} {\cal B}^\star \right)_1 \left( {\cal B} {\cal B}^\star \right)_2 
|\Psi^{(I)} \Psi^{(J)} \Phi^{(K)} \Phi^{(L)} \Phi^{(H)} \rangle ,
\label{defIJ} \ee
where the two states $\Psi$ are inserted at the punctures $I$ and $J$ and the 
states $\Phi$ are inserted at the other three punctures $K$, $L$ and $H$, we
find
\be
\boxed{\kappa^2 V_{\phi^3 \psi^2} = \frac{1}{\pi^2} \int_{{\cal A}_5} 
\left( F_{\phi^3 \psi^2}^{(12)} + F_{\phi^3 \psi^2}^{(15)} + F_{\phi^3 \psi^2}^{(25)} + 
F_{\phi^3 \psi^2}^{(34)} + 2 F_{\phi^3 \psi^2}^{(13)}
+ 2 F_{\phi^3 \psi^2}^{(23)} + 2 F_{\phi^3 \psi^2}^{(53)} 
\right) dx_1 dy_1 dx_2 dy_2} .
\label{phi3psi2}
\ee

\subsection{Contact terms of tachyons and dilatons}
We now list the results for the functions $F$ that we used in this
paper. The results of the integrations are shown in Tables
\ref{margCTable}, \ref{EffPot5ContactTable} and \ref{T4MTable}. We
start with the terms with tachyons and dilatons. The five-tachyon and
five-dilaton terms were calculated in \cite{quintic} so we don't
repeat them here. We will need the following open ghost correlators
\begin{align}
& A_{IJ} \equiv \langle (c_{-1} c_1)^{(I)}, c_{-1}^{(J)} \rangle_o \ , \quad 
B_{IJ} \equiv \langle (c_{-1} c_1)^{(I)}, c_1^{(J)} \rangle_o \label{correlators} \\
& C_{IJK} \equiv \langle c_1^{(I)}, c_1^{(J)}, c_1^{(K)} \rangle_o \ , \quad 
D_{IJK} \equiv \langle c_{-1}^{(I)}, c_1^{(J)}, c_1^{(K)} \rangle_o \ , \quad 
E_{IJK} \equiv \langle c_{-1}^{(I)}, c_{-1}^{(J)}, c_1^{(K)} \rangle_o .
\nonumber
\end{align}
Expressed in terms of the coefficients in the maps expansions
(\ref{mapsI}) and (\ref{maps5}), these are (defining $z_{IJ} \equiv
z_I-z_J$ and $\epsilon_I \equiv 8 \beta_I^2-6 \gamma_I$)
\begin{align}
& A_{IJ} = \rho_J \left( \beta_J - \beta_I - 2 \beta_I \beta_J z_{IJ} 
+ \frac{1}{2} \epsilon_J z_{IJ} (1-\beta_I z_{IJ}) \right) , \quad 
A_{5J} = \rho_J \left( \frac{1}{2} \epsilon_J (\beta_5 + z_J) - \beta_J \right) \nonumber \\
&B_{IJ} = \frac{1}{\rho_J} z_{IJ} (1-\beta_I z_{IJ}) , \quad 
B_{I5} = \frac{\beta_I}{\rho_5} , \quad B_{5J} = \frac{1}{\rho_J} (z_J + \beta_5) \nonumber \\
&C_{IJK} = \frac{1}{\rho_I \rho_J \rho_K} z_{IJ} z_{IK} z_{JK} , \quad 
C_{IJ5} = \frac{z_{JI}}{\rho_I \rho_J \rho_5} \nonumber \\
&D_{IJK} = \frac{\rho_I}{\rho_J \rho_K} \left( z_{JK} - 
\beta_I (z_{IK} + z_{IJ}) z_{JK} + 
\frac{1}{2} \epsilon_I z_{IJ} z_{JK} z_{IK} \right), \quad 
D_{IJ5} = \frac{\rho_I}{\rho_J \rho_5} \left( \beta_I - \frac{1}{2} \epsilon_I z_{IJ}
\right), \nonumber \\
& \hspace{24pt} D_{5IJ} = \frac{\rho_5}{\rho_I \rho_J} z_{JI} \left( 
z_I z_J + \beta_5 (z_I + z_J) + \frac{\epsilon_5}{2} \right)\nonumber \\
&E_{IJK} = \frac{\rho_I \rho_J}{\rho_K} \left( \beta_I - \beta_J + 
\beta_I \beta_J (z_{IJ} + z_{IK} - z_{JK}) + \frac{1}{2} \epsilon_J z_{JK} - 
\frac{1}{2} \beta_I \epsilon_J (z_{IJ} + z_{IK}) z_{JK} + \right. \nonumber \\
& \hspace{1.2cm} \left. + \frac{1}{2} \epsilon_I \left(-z_{IK} + \beta_J z_{IK} (z_{JK} - z_{IJ}) 
+ \frac{1}{2} \epsilon_J z_{IJ} z_{IK} z_{JK} \right) \right)
, \nonumber \\
& \hspace{24pt} E_{IJ5} = \frac{\rho_I \rho_J}{\rho_5} \left( \frac{1}{2} \beta_I \epsilon_J - 
\frac{1}{2} \epsilon_I \beta_J - \frac{1}{4} \epsilon_I \epsilon_J z_{IJ} \right),
\label{corrnot5} \end{align}
and it is understood that $I,J,K \neq 5$. We can now present the
results for the closed correlators.

\paragraph{Four tachyons and one dilaton}

\begin{align}
F_{t^4d}^{(3)} &= 
4 \Re \left( \frac{C_{1,1}^3}{\rho_1^2 \rho_2^2 \rho_3 \rho_4^2 \rho_5^2} \right)
\\
F_{t^4d}^{(I)} &= 
4 \Re \left(\frac{C_{IJK}}{\rho_3 \rho_4} \left( 
\frac{C_{2,1}^I \overline{C_{J4K}}}{\rho_3} + 
\frac{C_{1,1}^I \overline{C_{J3K}}}{\rho_4} \right) \right)
, \quad I \neq J \neq K \neq 3 \neq 4 .
\label{com1} \end{align}
Note that we are giving a transitive meaning to the inequality sign. So for example, 
by $\quad I \neq J \neq K \neq 3 \neq 4$ we really mean that $I$, $J$ and $K$ are pairwise 
distinct and that none of them is equal to $3$ or $4$. For a given $I$, there are two 
possible choices of $J$ and $K$ in equation (\ref{com1}), but they give the same result 
because the right-hand side of (\ref{com1}) is manifestly invariant under 
$J \leftrightarrow K$.

\paragraph{Three tachyons and two dilatons}

\begin{align}
F_{t^3d^2}^{(34)} =& 4 \frac{\left|C_{125}\right|^2}{\rho_3 \rho_4} \Re \left\{
C_{1,1}^3 C_{2,1}^4 - C_{1,1}^4 C_{2,1}^3 + C_{1,1}^3 \overline{C_{2,1}^4} - 
\overline{B_{1,1}^4} B_{2,1}^3 \right\}
\\
F_{t^3d^2}^{(I3)} =& 2 \Re \sum_{\genfrac{}{}{0pt}{}{J,K}{J \neq K \neq I \neq 3 \neq 4}} 
\left\{ C_{IJK} \left( \frac{\overline{C_{J4K}}}{\rho_3 \rho_4} 
\left( C_{1,1}^3 C_{2,1}^I - C_{1,1}^I C_{2,1}^3 + 
\overline{C_{1,1}^3} C_{2,1}^I - B_{1,1}^I \overline{B_{2,1}^3} \right) \right. \right.
\nonumber \\ 
& \left. \left. + \frac{\overline{C_{J3K}}}{\rho_4^2} \left( C_{1,1}^I \overline{C_{1,1}^3} - 
\overline{B_{1,1}^3} B_{1,1}^I \right) - 
\frac{\overline{D_{3JK}}}{\rho_3 \rho_4^2} B_{1,1}^I \right) \right\} , \quad I \neq 3,4
\\
F_{t^3d^2}^{(JK)} =& 4 \Re \left\{
\frac{1}{\rho_3 \rho_4} \left(C_{1,1}^J C_{2,1}^K - C_{1,1}^KC_{2,1}^J \right)
C_{IJK} \overline{C_{I34}} + \frac{1}{\rho_3^2 \rho_4^2} \overline{B_{KI}} B_{JI} 
\right. \nonumber \\ 
& - \frac{1}{\rho_3^2 \rho_4} \left( \overline{B_{2,1}^K} B_{JI} \overline{C_{I4K}} 
- B_{2,1}^J C_{IJ4} \overline{B_{KI}} \right) + 
\frac{1}{\rho_3 \rho_4^2} \left( B_{1,1}^J C_{IJ3} \overline{B_{KI}} - 
\overline{B_{1,1}^K} B_{JI} \overline{C_{I3K}} \right)
\nonumber \\
& - \frac{1}{\rho_3^2} \left( B_{2,1}^J \overline{B_{2,1}^K} - 
\overline{C_{2,1}^K} C_{2,1}^J \right) C_{IJ4} \overline{C_{I4K}} 
- \frac{1}{\rho_4^2} \left( B_{1,1}^J \overline{B_{1,1}^K} - 
\overline{C_{1,1}^K} C_{1,1}^J \right) C_{IJ3} \overline{C_{I3K}} 
\nonumber \\
& - \frac{1}{\rho_3 \rho_4} \left( \overline{B_{1,1}^K} B_{2,1}^J - 
C_{1,1}^J \overline{C_{2,1}^K} \right) C_{IJ4} \overline{C_{I3K}}
\nonumber \\
& \left. - \frac{1}{\rho_3 \rho_4} 
\left(B_{1,1}^J \overline{B_{2,1}^K} - \overline{C_{1,1}^K} C_{2,1}^J \right) 
C_{IJ3} \overline{C_{I4K}} \right\} , \quad I \neq J \neq K \neq 3 \neq 4 .
\end{align}

\paragraph{Two tachyons and three dilatons}

\begin{align}
F_{t^2d^3}^{(34)} =& 4 \Re \sum_{I \neq J \neq K \neq 3 \neq 4} \left\{ 
\frac{1}{\rho_3 \rho_4} C_{1,1}^I \left( \overline{B_{2,1}^J} B_{KI} 
\overline{C_{J34}} - B_{2,1}^J \overline{B_{K3}} C_{IJ4} \right) \right.
\nonumber \\
& + \frac{1}{\rho_3 \rho_4} C_{2,1}^I \left( \overline{B_{1,1}^J} B_{KI} 
\overline{C_{J43}} - B_{1,1}^J \overline{B_{K4}} C_{IJ3} \right)
+ \frac{1}{\rho_3} C_{IJ4} \overline{C_{34K}} \left( C_{1,1}^I M_2^{JK} + 
C_{2,1}^J B_{2,1}^I \overline{B_{1,1}^K} \right)
\nonumber \\
& + \frac{1}{\rho_4} C_{IJ3} \overline{C_{43K}} \left( C_{2,1}^I M_1^{JK} + 
C_{1,1}^J B_{1,1}^I \overline{B_{2,1}^K} \right) - 
\frac{1}{\rho_3 \rho_4}\left(\frac{1}{\rho_4} C_{1,1}^I B_{JI} 
\overline{B_{K3}} + \frac{1}{\rho_3} C_{2,1}^I B_{JI} \overline{B_{K4}} \right)
\nonumber \\
& \left. + \frac{1}{\rho_4^2} B_{1,1}^I C_{1,1}^J C_{IJ3} \overline{B_{K3}} + 
\frac{1}{\rho_3^2} B_{2,1}^I C_{2,1}^J C_{IJ4} \overline{B_{K4}} 
\right\}
\\
F_{t^2d^3}^{(JK)} =& 4 \Re \left\{ \overline{C_{IJK}} \left\{ 
\frac{1}{\rho_3 \rho_4} \left( C_{1,1}^4 \overline{B_{2,1}^I} - 
C_{2,1}^4 \overline{B_{1,1}^I} \right) D_{3JK} - 
\frac{1}{\rho_3 \rho_4} \left( C_{1,1}^3 \overline{B_{2,1}^I} - 
C_{2,1}^3 \overline{B_{1,1}^I} \right) D_{4JK} \right. \right.
\nonumber \\
& + \frac{1}{\rho_3} \left( C_{1,1}^4 \left(B_{2,1}^3 \overline{B_{2,1}^I} - 
C_{2,1}^3 \overline{C_{2,1}^I} \right) - C_{1,1}^3 \left( 
B_{2,1}^4 \overline{B_{2,1}^I} - C_{2,1}^4 \overline{C_{2,1}^I} \right) \right.
\nonumber \\
& \hspace{36pt} \left. - 
\overline{B_{1,1}^I} \left( B_{2,1}^3 C_{2,1}^4 - B_{2,1}^4 C_{2,1}^3 
\right) \right) C_{J4K}
\nonumber \\
& + \frac{1}{\rho_4} \left( C_{2,1}^3 \left(B_{1,1}^4 \overline{B_{1,1}^I} - 
C_{1,1}^4 \overline{C_{1,1}^I} \right) - C_{2,1}^4 \left( 
B_{1,1}^3 \overline{B_{1,1}^I} - C_{1,1}^3 \overline{C_{1,1}^I} \right) \right.
\nonumber \\
& \hspace{36pt} \left. - 
\overline{B_{2,1}^I} \left( B_{1,1}^4 C_{1,1}^3 - B_{1,1}^3 C_{1,1}^4 
\right) \right) C_{J3K}
\nonumber \\
& + \frac{1}{\rho_3} \left( \overline{C_{1,1}^I} \left(B_{2,1}^4 \overline{B_{2,1}^3} - 
C_{2,1}^4 \overline{C_{2,1}^3} \right) - \overline{C_{1,1}^3} \left( 
B_{2,1}^4 \overline{B_{2,1}^I} - C_{2,1}^4 \overline{C_{2,1}^I} \right) \right.
\nonumber \\
& \hspace{36pt} \left. - 
B_{1,1}^4 \left( \overline{B_{2,1}^3 C_{2,1}^I} - \overline{B_{2,1}^I C_{2,1}^3} 
\right) \right) C_{J4K}
\nonumber \\
& + \frac{1}{\rho_4} \left( \overline{C_{2,1}^I} \left(B_{1,1}^3 \overline{B_{1,1}^4} - 
C_{1,1}^3 \overline{C_{1,1}^4} \right) - \overline{C_{2,1}^4} \left( 
B_{1,1}^3 \overline{B_{1,1}^I} - C_{1,1}^3 \overline{C_{1,1}^I} \right) \right.
\nonumber \\
& \hspace{36pt} \left. - 
B_{2,1}^3 \left( \overline{B_{1,1}^4 C_{1,1}^I} - \overline{B_{1,1}^I C_{1,1}^4} 
\right) \right) C_{J3K} + \frac{1}{\rho_3 \rho_4} \left( \overline{C_{1,1}^I B_{2,1}^3} - 
\overline{C_{1,1}^3 B_{2,1}^I} \right) D_{4JK}
\nonumber \\
& \left. \left. + \frac{1}{\rho_3 \rho_4}
\left( \overline{C_{2,1}^I B_{1,1}^4} - \overline{C_{2,1}^4 B_{1,1}^I} \right) 
D_{3JK} \right\} \right\} , \quad I \neq J \neq K \neq 3 \neq 4 
\\
F_{t^2d^3}^{(I3)} =& 4 \Re \Biggl\{ \sum_{\genfrac{}{}{0pt}{}{J,K,L}{J \neq K \neq L \neq 3 \neq I}} 
\biggl\{ \frac{1}{\rho_3} C_{JKI} \overline{C_{3IL}} \left( C_{1,1}^J M_2^{KL} + 
C_{2,1}^K B_{2,1}^J \overline{B_{1,1}^L} \right) + 
\frac{1}{\rho_3^2} B_{2,1}^J C_{2,1}^K C_{JKI} \overline{B_{LI}} \biggr\} 
\nonumber \\
& + \frac{1}{\rho_4} 
\sum_{\genfrac{}{}{0pt}{}{J,K}{J \neq K \neq I \neq 3 \neq 4}} \biggl\{
\left( C_{2,1}^J \left(B_{1,1}^4 \overline{B_{1,1}^K} - 
C_{1,1}^4 \overline{C_{1,1}^K} \right) - 
C_{2,1}^4 \left( B_{1,1}^J \overline{B_{1,1}^K} - C_{1,1}^J \overline{C_{1,1}^K} 
\right) \right. \nonumber \\
& \hspace{12pt} \left.
- \overline{B_{2,1}^K} \left( B_{1,1}^4 C_{1,1}^J - B_{1,1}^J C_{1,1}^4 \right)
\right) C_{IJ3} \overline{C_{KI3}} + 
\frac{1}{\rho_3^2} \left( C_{2,1}^4 B_{JI} + C_{2,1}^J D_{4IJ} \right) \overline{B_{KI}}
\nonumber \\
& \hspace{12pt} + \frac{1}{\rho_3}
\left( B_{1,1}^4 C_{2,1}^K - B_{1,1}^K C_{2,1}^4 \right) 
C_{I3K} \overline{B_{JI}} + \frac{1}{\rho_3}
\left( B_{1,1}^J \overline{C_{2,1}^4} - B_{2,1}^J \overline{C_{1,1}^4} \right)
C_{IJ3} \overline{B_{KI}} \nonumber \\
& \hspace{12pt} + 
\frac{1}{\rho_3} \left( B_{1,1}^J \overline{C_{2,1}^K} - B_{2,1}^J 
\overline{C_{1,1}^K} \right)
C_{IJ3} \overline{D_{4IK}} + \frac{1}{\rho_3} C_{1,1}^J B_{2,1}^K
C_{IJK} \overline{D_{4I3}} \biggr\} \Biggr\} , \quad I \neq 3,4 .
\end{align}

\paragraph{One tachyon and four dilatons}

\begin{align}
F_{td^4}^{(3)} =& 4 \Re \Biggl\{ \sum_{I \neq J \neq K \neq L \neq 3} \left\{ 
\frac{1}{2\rho_3^2} M_2^{IJ} B_{KI} \overline{B_{LJ}} + 
\frac{1}{\rho_3} \left( C_{1,1}^I \overline{B_{2,1}^J C_{2,1}^K} + 
\overline{B_{1,1}^J} M_2^{IK} \right) B_{LI} \overline{C_{J3K}} \right.
\nonumber \\
& \hspace{12pt} \left. - \frac{1}{\rho_3} C_{1,1}^I B_{2,1}^J C_{2,1}^K 
\overline{B_{L3}} C_{IJK} + 
B_{1,1}^I C_{1,1}^J \overline{B_{2,1}^K C_{2,1}^L} C_{IJ3} \overline{C_{3KL}} 
+ \frac{1}{2} M_1^{IJ} M_2^{KL} C_{I3K} \overline{C_{J3L}} \right\}
\nonumber \\
& + \sum_{I \neq J \neq K \neq 3 \neq 4} \left\{ 
-\frac{1}{\rho_3^2 \rho_4} \overline{B_{2,1}^I} A_{J4} \overline{B_{KI}} 
+ \frac{1}{\rho_3 \rho_4} \left( C_{1,1}^I \overline{C_{2,1}^J} - 
\overline{B_{1,1}^J} B_{2,1}^I \right) B_{KI} \overline{D_{4J3}}
\right.
\nonumber \\
& \hspace{12pt} + \frac{1}{\rho_3 \rho_4} \left( \left( 
\overline{B_{1,1}^4} B_{2,1}^J + 
C_{1,1}^4 C_{2,1}^J - C_{1,1}^J C_{2,1}^4 - C_{1,1}^J \overline{C_{2,1}^4} 
\right) B_{IJ} \overline{B_{K3}} - 
C_{1,1}^I C_{2,1}^J \overline{B_{K3}} D_{4IJ} \right)
\nonumber \\
& \hspace{12pt} + \frac{1}{\rho_3 \rho_4} \overline{B_{1,1}^I B_{2,1}^J} A_{K4} 
\overline{C_{IJ3}} + \frac{1}{\rho_4} B_{1,1}^I C_{1,1}^J C_{IJ3} 
\left( \overline{C_{2,1}^K D_{43K}} + \overline{C_{2,1}^4 B_{K3}} \right)
\nonumber \\
& \hspace{12pt} + \frac{1}{\rho_4} \left( B_{1,1}^I C_{1,1}^J C_{2,1}^4 - 
B_{1,1}^I C_{1,1}^4 C_{2,1}^J + B_{1,1}^4 C_{1,1}^I C_{2,1}^J \right) 
C_{IJ3} \overline{B_{K3}} 
\nonumber \\
& \hspace{12pt} \left. + \frac{1}{\rho_4} \left(-M_1^{IJ} \overline{B_{2,1}^K} D_{4I3} 
\overline{C_{J3K}} + M_1^{4J} \overline{B_{2,1}^K} B_{I3} 
\overline{C_{J3K}} \right) \right\} \Biggr\}
\\
F_{td^4}^{(I)} =& 4 \Re \Biggl\{ \frac{1}{\rho_4} 
\sum_{\genfrac{}{}{0pt}{}{J,K,L}{J \neq K \neq L \neq I \neq 4}}
\biggl\{ B_{1,1}^J C_{11}^K C_{JKI} \left( C_{2,1}^L \overline{D_{4IL}} 
+ \overline{C_{2,1}^4 B_{LI}} \right) - M_1^{JK} \overline{B_{2,1}^L} 
D_{4JI} \overline{C_{KIL}}
\nonumber \\
& \hspace{12pt} + M_1^{4K} \overline{B_{2,1}^L} B_{JI} \overline{C_{KIL}} + 
\left( B_{1,1}^J C_{1,1}^K C_{2,1}^4 - B_{1,1}^J C_{1,1}^4 C_{2,1}^K + 
B_{1,1}^4 C_{1,1}^J C_{2,1}^K \right) C_{JKI} \overline{B_{LI}} \biggr\}
\nonumber \\
& + \sum_{\genfrac{}{}{0pt}{}{J,K,L,H}{J \neq K \neq L \neq H \neq I}}
\left\{ B_{1,1}^J C_{1,1}^K \overline{B_{2,1}^L C_{2,1}^H} C_{JKI} 
\overline{C_{ILH}} + \frac{1}{2} M_1^{JK} M_2^{LH} C_{JIL} 
\overline{C_{KIH}} \right\}
\nonumber\\
& + \frac{1}{\rho_3 \rho_4} 
\sum_{\genfrac{}{}{0pt}{}{J,K}{J \neq K \neq I \neq 3 \neq 4}} \biggl\{
\left( C_{1,1}^J \overline{C_{2,1}^K} - B_{2,1}^J \overline{B_{1,1}^K} 
\right) D_{3IJ} \overline{D_{4IK}} + \overline{B_{1,1}^J} B_{2,1}^K E_{34I} 
\overline{C_{IJK}}
\nonumber \\
& \hspace{12pt} + \left( \overline{C_{1,1}^3} C_{2,1}^K - 
\overline{B_{2,1}^3} B_{1,1}^K + C_{1,1}^3 C_{2,1}^K - C_{2,1}^3 C_{1,1}^K \right)
D_{4IK} \overline{B_{JI}} 
\nonumber \\
& \hspace{12pt} + \left( \overline{C_{2,1}^4} C_{1,1}^K - 
\overline{B_{1,1}^4} B_{2,1}^K + C_{2,1}^4 C_{1,1}^K - C_{1,1}^4 C_{2,1}^K \right)
D_{3IK} \overline{B_{JI}}
\nonumber \\
& \hspace{12pt} + \left(\overline{C_{2,1}^4} C_{1,1}^3  - \overline{B_{1,1}^4} 
B_{2,1}^3 + C_{2,1}^4 C_{1,1}^3 - C_{1,1}^4 C_{2,1}^3 \right) B_{JI} 
\overline{B_{KI}} \biggr\}
\nonumber \\
& + \frac{1}{\rho_3} 
\sum_{\genfrac{}{}{0pt}{}{J,K,L}{J \neq K \neq L \neq I \neq 3}} 
C_{IKL} \biggl\{
\left( \overline{C_{1,1}^J} B_{2,1}^K C_{2,1}^L + B_{1,1}^K 
\overline{M_2^{JL}} \right) \overline{D_{3IJ}} 
+ \Bigl( \overline{C_{1,1}^3} B_{2,1}^K C_{2,1}^L + 
B_{1,1}^K \overline{M_2^{3L}}
\nonumber \\
& \hspace{12pt} + C_{1,1}^3 B_{2,1}^K C_{2,1}^L - 
C_{1,1}^K B_{2,1}^3 C_{2,1}^L - C_{1,1}^L B_{2,1}^K C_{2,1}^3 \Bigr)
\overline{B_{JI}} \biggr\} \Biggr\} , \quad I \neq 3,4
\end{align}
The results of the integrations are shown in Table \ref{EffPot5ContactTable}.

\subsection{Contact terms of dilatons and marginal fields}

Since the ghost part of the marginal state $|A\rangle$, defined in
(\ref{marginal}), is that of a tachyon, and because the correlators
factorize into ghost and matter parts, we can recycle the results of
correlators of tachyons and dilatons. We just need to calculate the
matter correlators.  For the correlators of two marginals and three
dilatons, we have
\be
F_{a^2d^3}^{(IJ)} = F_{t^2d^3}^{(IJ)} \left| 
\langle \langle \alpha_{-1}^{(I)} \alpha_{-1}^{(J)} \rangle \rangle_o
\right|^2 , \quad I,J = 1, \ldots,5 ,
\ee
where the open matter correlators $\langle \langle \ldots \rangle \rangle_o$ are 
\be
\langle \langle \alpha_{-1}^{(I)} \alpha_{-1}^{(J)} \rangle \rangle_o = 
\frac{\rho_I \rho_J}{z_{JI}} , \quad 
\langle \langle \alpha_{-1}^{(I)} \alpha_{-1}^{(5)} \rangle \rangle_o = 
-\rho_I \rho_5 , \quad I,J = 1,\ldots,4 .
\ee
And for the correlators of four marginal fields and one dilaton we have 
\be
F_{a^4d}^{(I)} = F_{t^4d}^{(I)} \left| 
\langle \langle \alpha_{-1}^{(J)} \alpha_{-1}^{(K)} \alpha_{-1}^{(L)} 
\alpha_{-1}^{(H)} \rangle \rangle_o 
\right|^2 , \quad I = 1, \ldots,5 , \quad J \neq K \neq L \neq H \neq I .
\ee
For these matter correlators we find 
\begin{align}
& \langle \langle \alpha_{-1}^{(I)} \alpha_{-1}^{(J)} \alpha_{-1}^{(K)} 
\alpha_{-1}^{(5)} \rangle \rangle_o = 
-\rho_I \rho_J \rho_K \rho_5 \left( \frac{1}{z_{IJ}^2} + \frac{1}{z_{IK}^2} + 
\frac{1}{z_{JK}^2} \right) ,\quad I,J,K \neq 5 \nonumber \\
& \langle \langle \alpha_{-1}^{(1)} \alpha_{-1}^{(2)} \alpha_{-1}^{(3)} 
\alpha_{-1}^{(4)} \rangle \rangle_o = 
\rho_1 \rho_2 \rho_3 \rho_4 \left( \frac{1}{(\xi_1-\xi_2)^2} + 
\frac{1}{\xi_1^2 (1-\xi_2)^2} + \frac{1}{\xi_2^2 (1-\xi_1)^2} \right) .
\end{align}
The results of the integrations are shown in Table \ref{margCTable}.

\subsection{Contact terms of four tachyons and one field of level four}

The level four fields $f_1$, $f_2$, $f_3$ and $g_1$ were defined in
(\ref{field4}).  We will need a few new open correlators. We define
\be
P_{IJKL} \equiv \langle b_{-2} c_1^{(I)}, c_1^{(J)}, c_1^{(K)}, c_1^{(L)} 
\rangle_o , \quad Q_3 \equiv \langle c_1^{(1)}, c_1^{(2)}, b_{-2}^{(3)}, 
c_1^{(4)}, c_1^{(5)} \rangle_o , \quad 
G_I \equiv \langle \langle L_{-2}^{(I)} \rangle \rangle_o.
\ee
Elementary calculations give the following expressions for the correlators that 
we need
\begin{align}
P_{IJK5} =& \frac{\rho_I}{\rho_J \rho_K \rho_5} z_{JK} \left( 
\frac{1}{z_{IJ}} + \frac{1}{z_{IK}} - 3 \beta_I \right) \nonumber \\
P_{51IJ} =& \frac{\rho_5}{\rho_1 \rho_I \rho_J} z_{IJ} 
\left( z_I z_J \left( 1+\xi_1+\xi_2+3 \beta_5 \right) - \xi_1 \xi_2 \right)
\nonumber \\
Q_3 =& \frac{\rho_3^2}{\rho_1 \rho_2 \rho_4 \rho_5} 
\frac{\xi_2 (1-\xi_2)}{\xi_1 (1-\xi_1) (\xi_1-\xi_2)} \nonumber \\
G_I =& \frac{13}{6} \rho_I^2 \left( 2 \beta_I^2-\epsilon_I \right) ,
\end{align}
where $I,J,K = 1,\ldots,4$. And for the functions to integrate we find
\begin{align}
F_{t^4f_1}^{(3)} =& 
\frac{2}{\rho_4^2} \left( \left|B_{1,1}^3\right|^2 - \left|C_{1,1}^3\right|^2 \right) 
\left|C_{125}\right|^2
\\
F_{t^4f_1}^{(I)} =& 
4 \Re \Biggl\{ \frac{\left|D_{IJK}\right|^2}{2 \rho_3^2 \rho_4^2} + 
\frac{B_{1,1}^I}{\rho_3 \rho_4^2} C_{J3K} \overline{D_{IJK}} + 
\frac{B_{2,1}^I}{\rho_3^2 \rho_4} C_{J4K}\overline{D_{IJK}} 
\nonumber \\
& + \frac{\left|C_{J4K}\right|^2}{2\rho_3^2} \left( \left|B_{2,1}^I\right|^2 - 
\left|C_{2,1}^I\right|^2 \right) + 
\frac{\left|C_{J3K}\right|^2}{2\rho_4^2} \left( \left|B_{1,1}^I\right|^2 - 
\left|C_{1,1}^I\right|^2 \right) \nonumber \\ 
& + \frac{1}{\rho_3 \rho_4} \left( B_{1,1}^I \overline{B_{2,1}^I} - 
\overline{C_{1,1}^I} C_{2,1}^I \right) C_{J3K} \overline{C_{J4K}}
\Biggr\} , \quad I \neq J \neq K \neq 3 \neq 4 
\\
F_{t^4f_2}^{(I)} =& \frac{2}{\rho_3^2 \rho_4^2} \left|G_I\right|^2 
\left|C_{125}\right|^2 , \quad I = 1, \ldots, 5
\\
F_{t^4f_3}^{(3)} =& \frac{4}{\rho_3 \rho_4^2} \left|C_{125}\right|^2 
\Re \left\{ G_3 \overline{B_{1,1}^3} \right\} 
\\
F_{t^4f_3}^{(I)} =& 4 \Re \left\{
\frac{G_I C_{IJK}}{\rho_3^2 \rho_4^2} \left( \overline{D_{IJK}} + 
\rho_3 \overline{B_{1,1}^I 
C_{J3K}} + \rho_4 \overline{B_{2,1}^I C_{J4K}} \right) \right\} 
, \quad I \neq J \neq K \neq 3 \neq 4 \nonumber \\
\\
F_{t^4g_1}^{(3)} =& 4 \Re \left\{ C_{125} \left( \frac{C_{1,2}^3}{\rho_3 \rho_4^2} 
\overline{P_{3125}} + \frac{C_{2,2}^3}{\rho_3^2 \rho_4} \overline{Q_3}
\right) \right\}
\\
F_{t^4g_1}^{(I)} =& 4 \Re \left\{ C_{IJK} \left( \frac{C_{1,2}^I}{\rho_3 \rho_4^2} 
\overline{P_{IJ3K}} + \frac{C_{2,2}^I}{\rho_3^2 \rho_4} \overline{P_{IJ4K}}
\right) \right\} , \quad I \neq J \neq K \neq 3 \neq 4 
\end{align}
The results of the integrations are shown in Table \ref{T4MTable}.



\begin{thebibliography}{99}
\small
  
\bibitem{CSFT}
  B.~Zwiebach,
  ``Closed string field theory: Quantum action and the B-V master equation,''
  Nucl.\ Phys.\ B {\bf 390}, 33 (1993)
  [arXiv:hep-th/9206084];
  T.~Kugo, H.~Kunitomo and K.~Suehiro,
  ``Nonpolynomial Closed String Field Theory,''
  Phys.\ Lett.\ B {\bf 226}, 48 (1989);
  T.~Kugo and K.~Suehiro,
  ``Nonpolynomial Closed String Field Theory: Action And Its Gauge Invariance,''
  Nucl.\ Phys.\ B {\bf 337}, 434 (1990);
  H.~Sonoda and B.~Zwiebach,
  ``Covariant Closed String Theory Cannot Be Cubic,''
  Nucl.\ Phys.\ B {\bf 336}, 185 (1990);
  B.~Zwiebach,
  ``Quantum Closed Strings From Minimal Area,''
  Mod.\ Phys.\ Lett.\ A {\bf 5}, 2753 (1990);
  M.~Kaku,
  ``Geometric Derivation Of String Field Theory From First Principles: Closed
  Strings And Modular Invariance,''
  Phys.\ Rev.\ D {\bf 38}, 3052 (1988);
  M.~Saadi and B.~Zwiebach,
  ``Closed String Field Theory From Polyhedra,''
  Annals Phys.\  {\bf 192}, 213 (1989);
  B.~Zwiebach,
  ``Consistency Of Closed String Polyhedra From Minimal Area,''
  Phys.\ Lett.\ B {\bf 241}, 343 (1990);
  B.~Zwiebach,
  ``How Covariant Closed String Theory Solves A Minimal Area Problem,''
  Commun.\ Math.\ Phys.\  {\bf 136}, 83 (1991).

\bibitem{Kostelecky:1990mi}
  V.~A.~Kostelecky and S.~Samuel,
  ``Collective physics in the closed bosonic string,''
  Phys.\ Rev.\ D {\bf 42} (1990) 1289.

\bibitem{Belo-Zwie}
  A.~Belopolsky and B.~Zwiebach,
  ``Off-shell closed string amplitudes: Towards a computation of the tachyon potential,''
  Nucl.\ Phys.\ B {\bf 442}, 494 (1995)
  [arXiv:hep-th/9409015].

\bibitem{Belo}
  A.~Belopolsky,
  ``Effective Tachyonic potential in closed string field theory,''
  Nucl.\ Phys.\ B {\bf 448}, 245 (1995)
  [arXiv:hep-th/9412106].

\bibitem{vacuum}
  H.~Yang and B.~Zwiebach,
  ``A closed string tachyon vacuum?,''
  JHEP {\bf 0509}, 054 (2005)
  [arXiv:hep-th/0506077].

\bibitem{quartic}
  N.~Moeller,
  ``Closed bosonic string field theory at quartic order,''
  JHEP {\bf 0411} (2004) 018
  [arXiv:hep-th/0408067].

\bibitem{Yang:2005rw}
  H.~Yang and B.~Zwiebach,
  ``Rolling closed string tachyons and the big crunch,''
  JHEP {\bf 0508}, 046 (2005)
  [arXiv:hep-th/0506076].

\bibitem{Moe-Yang}
  N.~Moeller and H.~Yang,
  ``The nonperturbative closed string tachyon vacuum to high level,''
  JHEP {\bf 0704}, 009 (2007)
  [arXiv:hep-th/0609208].

\bibitem{quintic}
  N.~Moeller,
  ``Closed bosonic string field theory at quintic order: Five-tachyon contact
  term and dilaton theorem,''
  JHEP {\bf 0703}, 043 (2007)
  [arXiv:hep-th/0609209].

\bibitem{marginal}
  H.~Yang and B.~Zwiebach,
  ``Testing closed string field theory with marginal fields,''
  JHEP {\bf 0506}, 038 (2005)
  [arXiv:hep-th/0501142].

\bibitem{dilaton}
  H.~Yang and B.~Zwiebach,
  ``Dilaton deformations in closed string field theory,''
  JHEP {\bf 0505}, 032 (2005)
  [arXiv:hep-th/0502161].

\bibitem{Taylor:2002fy}
  W.~Taylor,
  ``A perturbative analysis of tachyon condensation,''
  JHEP {\bf 0303} (2003) 029
  [arXiv:hep-th/0208149].

\bibitem{BST}
  M.~Beccaria and C.~Rampino,
  ``Level truncation and the quartic tachyon coupling,''
  JHEP {\bf 0310} (2003) 047
  [arXiv:hep-th/0308059].

\bibitem{twisted}
  Y.~Okawa and B.~Zwiebach,
  ``Twisted tachyon condensation in closed string field theory,''
  JHEP {\bf 0403}, 056 (2004)
  [arXiv:hep-th/0403051].
  
\bibitem{cyclic}
  P.~J.~Steinhardt and N.~Turok,
  ``The cyclic model simplified,''
  New Astron.\ Rev.\  {\bf 49}, 43 (2005)
  [arXiv:astro-ph/0404480];
  J.~Khoury,
  ``A briefing on the ekpyrotic / cyclic universe,''
  arXiv:astro-ph/0401579;
  J.~Khoury, P.~J.~Steinhardt and N.~Turok,
  ``Designing cyclic universe models,''
  Phys.\ Rev.\ Lett.\  {\bf 92}, 031302 (2004)
  [arXiv:hep-th/0307132];
  J.~Khoury, B.~A.~Ovrut, N.~Seiberg, P.~J.~Steinhardt and N.~Turok,
  ``From big crunch to big bang,''
  Phys.\ Rev.\  D {\bf 65}, 086007 (2002)
  [arXiv:hep-th/0108187].

\end{thebibliography}
\end{document}